\documentclass[AMA,STIX1COL]{WileyNJD-v2}

\articletype{Article Type}%

\received{Date Month Year}
\revised{Date Month Year}
\accepted{Date Month Year}

\copyyear{2023}
\startpage{1}

\usepackage{ifpdf}
\usepackage{color, colortbl}
\usepackage{tikz}
\usepackage{array}
\usepackage{multirow}
\usepackage{listings}
\usepackage{xcolor}
\usepackage{amsmath}
\usepackage{graphicx}
\usepackage{subcaption}
\usepackage{url}
\usepackage{booktabs}
\usepackage{pifont}
\usepackage{adjustbox,siunitx}

\lstdefinelanguage{json}{
    basicstyle=\normalfont\ttfamily,
    stringstyle=\color{darkgreen},
    literate=
     *{0}{{{\color{numb}0}}}{1}
      {:}{{{\color{punct}{:}}}}{1},
    morestring=[b]"
}
\colorlet{punct}{red!60!black}
\definecolor{background}{HTML}{EEEEEE}
\definecolor{darkgreen}{HTML}{006400}
\definecolor{delim}{RGB}{20,105,176}
\colorlet{numb}{magenta!60!black}

\begin{document}

\title{CloudNativeSim: a toolkit for modeling and simulation of cloud-native applications}

\author[1,2]{Jingfeng Wu}

\author[1]{Minxian Xu}

\author[1,3]{Yiyuan He}

\author[1]{Kejiang Ye}

\author[4]{Chengzhong Xu}

\authormark{Wu \textsc{et al.}}
\titlemark{CloudNativeSim: a toolkit for modeling and simulation of cloud-native applications}

\address[1]{\orgdiv{Shenzhen Institutes of Advanced Technology}, \orgname{Chinese Academy of Sciences}, \orgaddress{\state{Shenzhen}, \country{China}}}

\address[2]{\orgdiv{University of Chinese Academy of Sciences}, \orgaddress{\country{China}}}

\address[3]{\orgdiv{Southern University of Science and Technology}, \orgaddress{\state{Shenzhen}, \country{China}}}

\address[4]{\orgdiv{State Key Lab of IOTSC, Department of Computer Science}, \orgname{University of Macau}, \orgaddress{\state{Macau SAR}, \country{China}l}}


\corres{Minxian Xu, Shenzhen Institutes of Advanced Technology, Chinese Academy of Sciences, China \\ \email{mx.xu@siat.ac.cn}}

\abstract[Abstract]{
Cloud-native applications are increasingly becoming popular in modern software design. Employing a microservice-based architecture into these applications is a prevalent strategy that enhances system availability and flexibility. However, cloud-native applications introduce new challenges, including frequent inter-service communication and the management of heterogeneous codebases and hardware, resulting in unpredictable complexity and dynamism. Furthermore, as applications scale, only limited research teams or enterprises possess the resources for large-scale deployment and testing, which impedes progress in the cloud-native domain. To address these challenges, we propose CloudNativeSim, a simulator for cloud-native applications with a microservice-based architecture. CloudNativeSim offers several key benefits: (i) comprehensive and dynamic modeling for cloud-native applications, (ii) an extended simulation framework with new policy interfaces for scheduling cloud-native applications, and (iii) support for customized application scenarios and user feedback based on Quality of Service (QoS) metrics. CloudNativeSim can be easily deployed on standard computers to manage a high volume of requests and services. Its performance was validated through a case study, demonstrating higher than 94.5\% accuracy in terms of response time simulation. The study further highlights the feasibility of CloudNativeSim by illustrating the effects of various scaling policies.
}


\maketitle

\section{Introduction}\label{sec:introduction}

\begin{figure}
    \centering
    \includegraphics[width=0.85\linewidth]{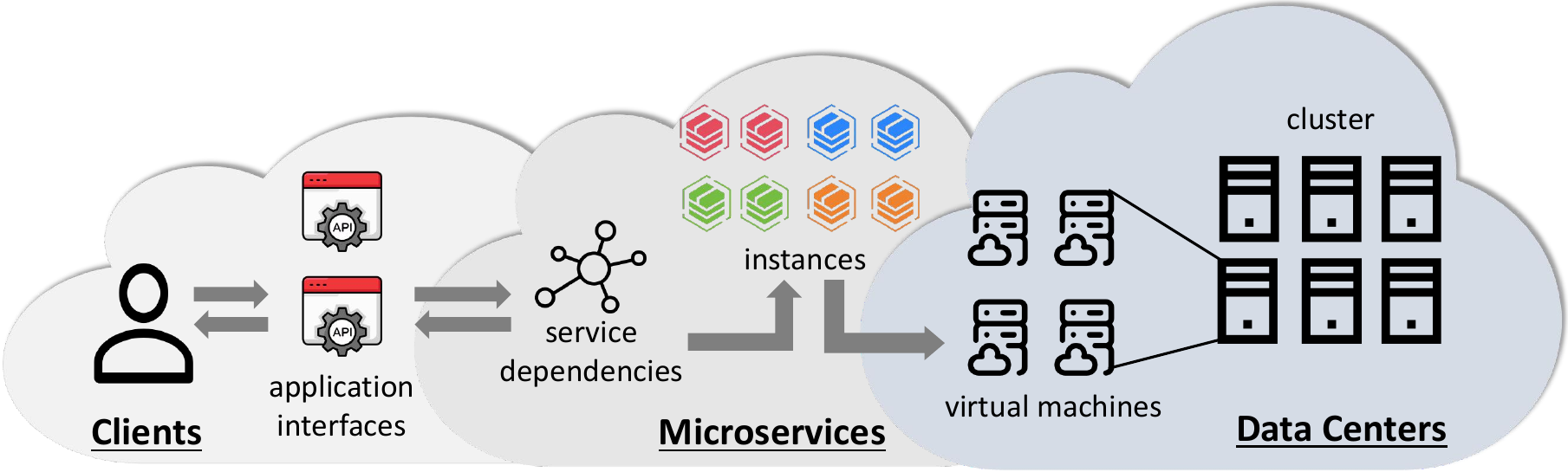}
    \caption{A demonstration of a cloud-native application architecture, showing the interaction between clients, application interfaces, microservices, and data centers.}
    \label{fig:introduction}
    \vspace{-1em}
\end{figure}

In modern software development, the trend of cloud-native applications is on the rise. More and more applications are migrated to cloud-native environments, offering reliable and flexible functions. When choosing the development architecture for cloud-native products, a microservice-based architecture is considered more efficient compared to traditional monolithic architectures \cite{Sylemez2023MicroserviceRA,enabler}. Figure \ref{fig:introduction} presents a demonstration of a cloud-native application based on microservice-based architecture, detailing the interactions between clients, microservices, and data centers. Clients send requests through application interfaces, which generate various service dependencies that are mapped to different instances (e.g., pods and containers). These microservices instances are deployed in suitable resource environments, supported by the resources of the data center, ensuring efficient and reliable service delivery. This architecture provides numerous advantages, including high availability and decentralization. These advantages can accelerate development and deployment processes, offer flexible auto-scaling mechanisms, and reduce maintenance costs \cite{exp-report,Xu2023PracticeOA,drpc}.

However, the transition to a microservice-based architecture faces challenges of \textbf{complexity and dynamism} \cite{heterogeneityAndDynamicity,transtion,applications-heterogeneity}. A microservice-based architecture typically operates on heterogeneous code and hardware. An application built on a microservice-based architecture may consist of intricate modules and complex dependencies. Moreover, to offer adaptable virtualization for these modules, microservice-based architecture introduces numerous entities, such as requests and services. As the application scales, the communications between these entities become more frequent, requiring the coordination of scheduling policies, and leading to an uncontrollable dynamism.

Furthermore, researchers face significant challenges in deploying large-scale cloud-native applications. Large-scale applications require a robust and extensive data center capable of handling numerous requests and allocating sufficient resources to internal entities. Few teams have the capability to fulfill these requirements, let alone to assess scheduling policies based on microservice-based architecture.

To address challenges such as microservice dependency modeling, resource scheduling under high dynamism, and limited resources for large-scale deployments, we propose CloudNativeSim, a toolkit for modeling and simulation of cloud-native applications, which possesses the following key \textbf{features}:
\begin{itemize}

\item \textbf{Comprehensive modeling approach}: CloudNativeSim is a comprehensive modeling framework designed specifically for cloud-native applications and microservice-based architectures. By incorporating essential components such as services, pods or containers, virtual machines (VMs), and datacenters, it offers more comprehensive object modeling capabilities compared to existing simulators.

\item \textbf{File registration mechanism}: To address the inherent complexity of microservice architectures, which involve numerous configuration units like services and pods, CloudNativeSim implements a file-based registration system. This mechanism enables batch registration of architectural components through configuration files, significantly reducing the manual configuration burden and potential for errors.

\item \textbf{Dynamic request generation and distribution}: CloudNativeSim features a built-in request generator that accurately simulates real-world scenarios. This component handles the dynamic process of clients randomly generating requests and intelligently distributing them across different services, allowing users to test various load patterns and service interactions.

\item \textbf{Innovative cloudlet scheduling mechanism}: Recognizing the limitations of traditional scheduling approaches in microservice environments, CloudNativeSim introduces an enhanced cloudlet scheduling mechanism. During runtime, cloudlets traverse through the service dependency graph under scheduler supervision, managing complex state transitions that include waiting, execution, and request propagation phases. This mechanism is specifically designed to handle the dynamic nature of microservice architectures, where requests flow through multiple interconnected services.

\item \textbf{Critical path-based latency calculation}: The simulator models service dependencies as a Directed Acyclic Graph (DAG), enabling detailed performance analysis. By identifying critical paths within parallel execution chains, CloudNativeSim provides accurate latency calculations that reflect real-world service interactions and potential bottlenecks.

\item \textbf{Dynamic QoS metrics record}: CloudNativeSim tracks and records dynamic changes in both system structure and QoS metrics during runtime. The QoS metrics include request-related measurements such as latency, Requests Per Second (RPS) for different Application Programming Interfaces (APIs), and Service Level Objective (SLO) violation rates as well as resource utilization metrics (CPU and memory usage) at both instance and VM levels. Through multiple channels including system logs and Grafana \footnote{https://grafana.com/} integration, these dynamic metrics can be monitored and visualized in real-time, providing comprehensive insights into system behavior and performance evolution.

\item \textbf{Extensible policy interfaces}: To accommodate the diverse requirements of microservice architectures, CloudNativeSim provides a comprehensive set of policy interfaces. These interfaces support various operational aspects including service allocation, scaling decisions, and resource management, allowing users to implement and evaluate customized policies tailored to their specific needs.

\end{itemize}

The key \textbf{contributions} of this work include: (i) a comprehensive modeling approach for cloud-native applications and microservice-based architecture, (ii) the ability to run on a standard personal computer while handling a large volume of requests and services concurrently, (iii) achieving a response time simulation accuracy of over 94.5\%, and (iv) providing a range of scheduling policy interfaces, enabling customized implementation and evaluation of various scheduling policies.

The structure of this paper is organized as follows: Section \ref{sec 2} presents a discussion of contemporary distributed system simulations and modeling, highlighting the unique capabilities of CloudNativeSim. Section \ref{sec 3} details CloudNativeSim's simulation framework, emphasizing the functions of each architectural component. Section \ref{sec 4} describes the key components related to requests and cloudlets, presenting advanced application modeling. Section \ref{sec 5} discusses the design of service scheduling policies, including resource allocation strategies and service scaling mechanisms. Finally, Section \ref{sec 6} provides comprehensive evaluations of CloudNativeSim, validating its simulation capacity and response time accuracy, and demonstrating the request generator and scaling algorithms. Section \ref{sec 7} concludes the paper with a brief summary and suggestions for future research directions.

\section{Related Work}\label{sec 2}

As cloud computing technology continues to advance, the number of cloud-native applications and the scale of microservice-based systems have significantly increased. To evaluate the effectiveness of scheduling algorithms in cloud environments, researchers have developed numerous simulators to support performance evaluations across various scenarios. These simulators provide a safe and controlled environment for performance evaluation and optimizing different scheduling policies. However, although recent studies have begun to model microservice-based architectures, existing simulators still lack adequate support for these architectures. For example, they often fail to accurately simulate the complex interactions and dependencies between microservices, which are crucial for evaluating the performance of scheduling algorithms. Additionally, existing simulators face limitations in scalability and configurability, making it challenging to meet the specific needs of different application scenarios. Therefore, developing more precise and flexible simulators has become a critical research direction to better support the practical application and optimization of microservice-based architectures in cloud computing environments.

Table \ref{table:comparison} presents a comparative overview of CloudNativeSim and other simulation tools, highlighting their capabilities in various environment simulations, including resource and data center management, service modeling, and task scheduling. This comparison aims to emphasize the strengths and limitations of each toolkit, providing insights into their suitability for different research and application scenarios in cloud computing.

\begingroup
    \renewcommand*{\arraystretch}{1.5}%
    \definecolor{tabred}{RGB}{230,36,0}%
    \definecolor{tabgreen}{RGB}{0,116,21}%
    \definecolor{taborange}{RGB}{255,124,0}%
    \definecolor{tabbrown}{RGB}{171,70,0}%
    \definecolor{tabyellow}{RGB}{255,253,169}%
    \newcommand*{\redtriangle}{\textcolor{tabred}{\ding{115}}}%
    \newcommand*{\greenbullet}{\textcolor{tabgreen}{\ding{108}}}%
    \newcommand*{\orangecirc}{\textcolor{taborange}{\ding{109}}}%
    \newcommand*{\headformat}[1]{{\small#1}}%
    \newcommand*{\vcorr}{%
      \vadjust{\vspace{-\dp\csname @arstrutbox\endcsname}}%
      \global\let\vcorr\relax
    }%

    \newcommand*{\HeadAux}[1]{%
      \multicolumn{1}{@{}r@{}}{%
        \vcorr
        \sbox0{\headformat{\strut #1}}%
        \sbox2{\headformat{Complex Data Movement}}%
        \sbox4{\kern\tabcolsep\redtriangle\kern\tabcolsep}%
        \sbox6{\rotatebox{25}{\rule{0pt}{\dimexpr\ht0+\dp0\relax}}}%
        \sbox0{\raisebox{.5\dimexpr\dp0-\ht0\relax}[0pt][0pt]{\unhcopy0}}%
        \kern7.5mm 
        \rlap{\raisebox{.25\wd4}{\rotatebox{25}{\unhcopy0}}}
        \kern.25\wd1 %
        \ifx\HeadLine Y%
            \dimen0=\dimexpr\wd2+.1\wd4\relax 
            \rlap{\rotatebox{25}{\hbox{\vrule width 2.5cm height .1pt}}}%
        \fi
      }%
    }%
    \newcommand*{\head}[1]{\HeadAux{\global\let\HeadLine=Y#1}}%
    \newcommand*{\headNoLine}[1]{\HeadAux{\global\let\HeadLine=N#1}}%
    \noindent

\begin{table*}[htbp]
        \begin{footnotesize}
        \begin{flushleft}
          \begin{tabular}{
                    >{\bfseries}lc|>{\quad}c
                    *{3}{c|}c|>{\quad}c
                    *{4}{c|}c|>{\quad}c
                    *{2}{c|}c%
                }%
                &
                \head{\footnotesize Simulator Names} &
                &
                \head{\footnotesize Execution Environment} &
                \head{\footnotesize Datacenter Configuration} &
                \head{\footnotesize Resource Characteristic} &
                \head{\footnotesize Instance Management} &
                &
                \head{\footnotesize ServiceChain Modeling} &
                \head{\footnotesize Service Allocation} &
                \head{\footnotesize Horizontal Scaling} &
                \head{\footnotesize Vertical Scaling} &
                \head{\footnotesize Service Migration} &
                &
                \head{\footnotesize Tasks Scheduling} &
                \head{\footnotesize Request Processing} & 
                \headNoLine{\footnotesize QoS Metrics} \\

                \sbox0{S}%
                \rule{0pt}{\dimexpr\ht0 + 2ex\relax}%
                CloudSim 3.0 \cite{cloudsim} & \textcolor{tabred} && 
                Cloud 
                & \checkmark & \checkmark &  && &  &  &  & & 
                & \checkmark &  &  \\
                \hline

                CloudSim Plus \cite{cloudsimplus} & \textcolor{tabred} && 
                Cloud 
                & \checkmark & \checkmark & \checkmark && &  &  &  & & 
                & \checkmark &  &  \\
                \hline
    
                NetworkCloudSim \cite{networkcloudsim}  & \textcolor{tabred}  &&
                Cloud & \checkmark & \checkmark &  &&
                  &  &  &  &&
                & \checkmark &  &   \\
                \hline
                
                iFogSim \cite{ifogsim,ifogsim2}  & \textcolor{tabred} &&
                Fog/Edge & \checkmark & \checkmark &  &&
                  & \checkmark & \checkmark &  &\checkmark&
                & \checkmark & \checkmark&  \checkmark \\
                \hline
                
                ContainerCloudSim \cite{containercloudsim}  & \textcolor{tabred}  &&
                Cloud & \checkmark & \checkmark & \checkmark &&
                  &  &  &  &&
                & \checkmark &  &   \\
                \hline
                
                ServiceSim \cite{servicesim} & \textcolor{tabred}   &&
                Cloud/Edge & \checkmark & \checkmark & \checkmark && \checkmark
                & \checkmark &  &  & \checkmark && 
                \checkmark & \checkmark &  \checkmark \\
                \hline
                
                ColocationSim \cite{colocationsim} & \textcolor{tabred}  &&
                Cloud & \checkmark & \checkmark & \checkmark &&
                &  & &  &  &&
                 & \checkmark &  \\
                \hline
    
                PerfSim \cite{perfsim} & \textcolor{tabred}  &&
                Cloud/Edge & \checkmark & \checkmark& &&
                \checkmark&  & \checkmark & \checkmark &  &&
                 & \checkmark &   \\
                \hline
                
    
                \rowcolor{tabyellow}%
                CloudNativeSim & \textcolor{tabred}  &&
                Cloud & \checkmark & \checkmark & \checkmark &&
                \checkmark & \checkmark & \checkmark & \checkmark & \checkmark &&
                \checkmark & \checkmark & \checkmark \\
            \end{tabular}%
        \end{flushleft}
        \end{footnotesize}
    \caption{A detailed comparison of CloudNativeSim with other contemporary simulators highlights their unique capabilities in service modeling, horizontal and vertical scaling, and other cloud environment simulations. This comparison emphasizes CloudNativeSim's advanced features, particularly its comprehensive support for microservice-based architecture simulations.}
    \label{table:comparison}
    \vspace{-1em}
\end{table*}
\endgroup

As cloud computing technology evolves, a growing array of simulators have been developed to support performance evaluations from cloud to edge environments. CloudSim \cite{cloudsim} and its derivatives \cite{cloudsimplus,networkcloudsim,ifogsim,ifogsim2,containercloudsim,servicesim} stand out as the principal tools in this domain. Based on a discrete-event simulation framework, CloudSim supports extensive data centers and energy-efficient computational resources. It introduces simulation events and entities, facilitating detailed modeling and analysis. CloudSim employs the concept of "cloudlets" to represent tasks, with their length indicating the number of instructions to be processed. These cloudlets can be managed by time-sharing or space-sharing methods to optimize the utilization of processing elements (PE) in virtual machines. After version 3.0, CloudSim has incorporated new features, including support for Software-Defined Networking (SDN) and containers. However, the fundamental cloudlet-based task model poses significant challenges for microservice simulation. Specifically, the basic structure of cloudlets lacks the flexibility to represent the dynamic nature of microservice interactions, service dependencies, and fine-grained resource requirements characteristic of microservice architectures. Given CloudSim's extensible architecture that supports the customization of new entities and events, CloudNativeSim and various other derivatives have been developed upon this foundation, each specializing in distinct aspects of cloud computing simulation.


CloudSim Plus \cite{cloudsimplus}, an iteration of CloudSim 3.0, is redesigned using Java 17 to reduce code duplication and promote code reuse, enhancing both scalability and precision. This iteration offers a cleaner, more functional, and user-friendly interface, and introduces more advanced functions such as auto-scaling for virtual machines, heuristics for resource management, and lifetime control of entities. Despite these improvements, CloudSim Plus inherits the fundamental limitations of CloudSim's cloudlet-based architecture. The auto-scaling mechanisms, while suitable for virtual machines, are not designed to handle the rapid scaling and complex dependencies of microservices.

Following CloudSim's foundational framework, numerous simulators have been developed. Among them, NetworkCloudSim \cite{networkcloudsim}, an extension of CloudSim, is specifically designed to simulate network communications in cloud computing environments. It incorporates a network model into CloudSim, allowing researchers and developers to account for factors such as bandwidth limitations and network latencies within and between data centers. This enhancement enables more precise simulation and evaluation of cloud computing applications' performance. NetworkCloudSim introduces a TaskStage field in the cloudlet and considers the derivative relationships between cloudlets, providing a foundation for modeling task dependencies. While these features are valuable for traditional distributed applications, the current implementation focuses primarily on network-level communication rather than service-level interactions that characterize microservice architectures.

iFogSim \cite{ifogsim}, a widely used extension of CloudSim, introduces a controller-centric framework that focuses on the mobility of Internet of Things (IoT) devices and clustering mechanisms. Its second version \cite{ifogsim2} extends to include microservice-based features, such as service discovery, placement, horizontal scaling, and load balancing. It also addresses the deployment and routing of microservices, which is crucial for system flexibility \cite{router}. However, being primarily designed for Fog or Edge environments, its architecture and resource management mechanisms are optimized for edge computing scenarios rather than cloud-native environments, where different constraints and requirements may apply.

ContainerCloudSim \cite{containercloudsim}, by introducing container technology, represents a significant innovation in CloudSim. It builds a detailed container abstraction on top of virtual machines and extends new strategies for container placement, selection, provisioning, and overbooking. Although ContainerCloudSim does not emphasize microservice-based architectures, its in-depth portrayal of containers provides a robust framework for modeling instances of services. However, extending ContainerCloudSim to support microservice architectures presents considerable challenges. First, its extensive code overrides of CloudSim's base implementation create architectural constraints. More importantly, its fundamental design follows a "VM-to-containers" one-to-many mapping paradigm, which makes it difficult to implement the many-to-many "services-to-instances" relationships characteristic of microservice architectures. Adapting ContainerCloudSim to support such relationships would require substantial modifications to its core architecture and significantly increase code complexity.

ServiceSim \cite{servicesim}, while representing an advancement in cloud simulation by integrating features from iFogSim and NetworkCloudSim, demonstrates the common limitations faced by current simulators. ServiceSim successfully implements essential functionalities such as service discovery and load balancing, while incorporating sophisticated task staging mechanisms. However, it still leaves gaps in comprehensive microservice modeling and policy interface flexibility. These limitations reflect the broader challenge of adapting existing simulation frameworks to fully support modern microservice architectures.

ColocationSim \cite{colocationsim} exemplifies an alternative approach to microservice simulation through queuing theory models. While innovative in its handling of latency-critical and best-effort jobs, this mathematical approach highlights the inherent difficulties in accurately modeling microservice behaviors. The simulator reliance on traditional queuing models, though computationally efficient, struggles to capture the dynamic nature and complex interdependencies characteristic of microservice architectures. This limitation underscores the broader challenge of finding an appropriate balance between model simplicity and accurate representation of microservice complexity.

PerfSim \cite{perfsim} offers valuable insights through its distinctive approach to cloud simulation. By incorporating discrete event simulation mechanisms, it excels in modeling and simulating complex service dependencies, with particular strength in analyzing their performance implications. The simulator demonstrates sophisticated capabilities in mathematical modeling of service dependencies and supports both horizontal and vertical scaling mechanisms for performance optimization. Its robust handling of user requests and network packet transmission simulation provides a solid foundation for performance analysis. However, PerfSim's primary focus on service chain performance modeling, rather than comprehensive cloud-native applications, limits its broader applicability. 

In summary, while various cloud computing simulators have emerged to address different aspects of cloud and edge computing environments, they exhibit distinct limitations in their ability to comprehensively simulate modern microservice architectures. Traditional simulators like iFogSim and NetworkCloudSim, while robust in their respective focused areas, present significant challenges when attempting substantial modifications to support microservice-specific features. While specialized microservice simulators such as ServiceSim and PerfSim exist, their architectural approaches differ fundamentally from what is needed to fully capture the nature of cloud-native applications. Current limitations in microservice architecture modeling include incomplete entity representation, simplified request arrival patterns, and basic service scheduling strategies.

To address these systemic limitations and advance the field of cloud-native simulation, we propose CloudNativeSim with the following design objectives: (i) \textbf{File Registration Mechanism}: To manage the increased complexity introduced by microservice architectures, which involve numerous units such as services and Pods, CloudNativeSim implements a file-based batch registration system that significantly simplifies configuration management. (ii) \textbf{Dynamic Request Generation and Distribution}: The simulator features a built-in request generator that accurately models real-world scenarios by simulating how clients dynamically generate and distribute requests across various services. (iii) \textbf{Enhanced Cloudlet Scheduling}: Recognizing the limitations of traditional cloudlet scheduling in handling microservice architectures, CloudNativeSim introduces an advanced scheduling mechanism specifically designed to address the dynamic and complex nature of these systems. (iv) \textbf{Delay Analysis and Critical Path Detection}: CloudNativeSim models service dependencies as Directed Acyclic Graphs (DAGs), enabling precise calculation of request delays and identification of critical paths within parallel service chains. (v) \textbf{Comprehensive QoS Metrics}: The simulator provides extensive quality-of-service monitoring capabilities, offering detailed request-related feedback through system logs and integration with Grafana for enhanced visualization and analysis. and (vi) \textbf{Lightweight and Extensible Policy Framework}: To accommodate the diverse requirements of microservice architectures, CloudNativeSim implements flexible and lightweight interfaces for various policies, including service allocation and scaling, ensuring adaptability to different deployment scenarios.

\section{Simulation Framework of CloudNativeSim}\label{sec 3}

\begin{figure}[htb] 
\centering 
\includegraphics[width=0.75\textwidth]{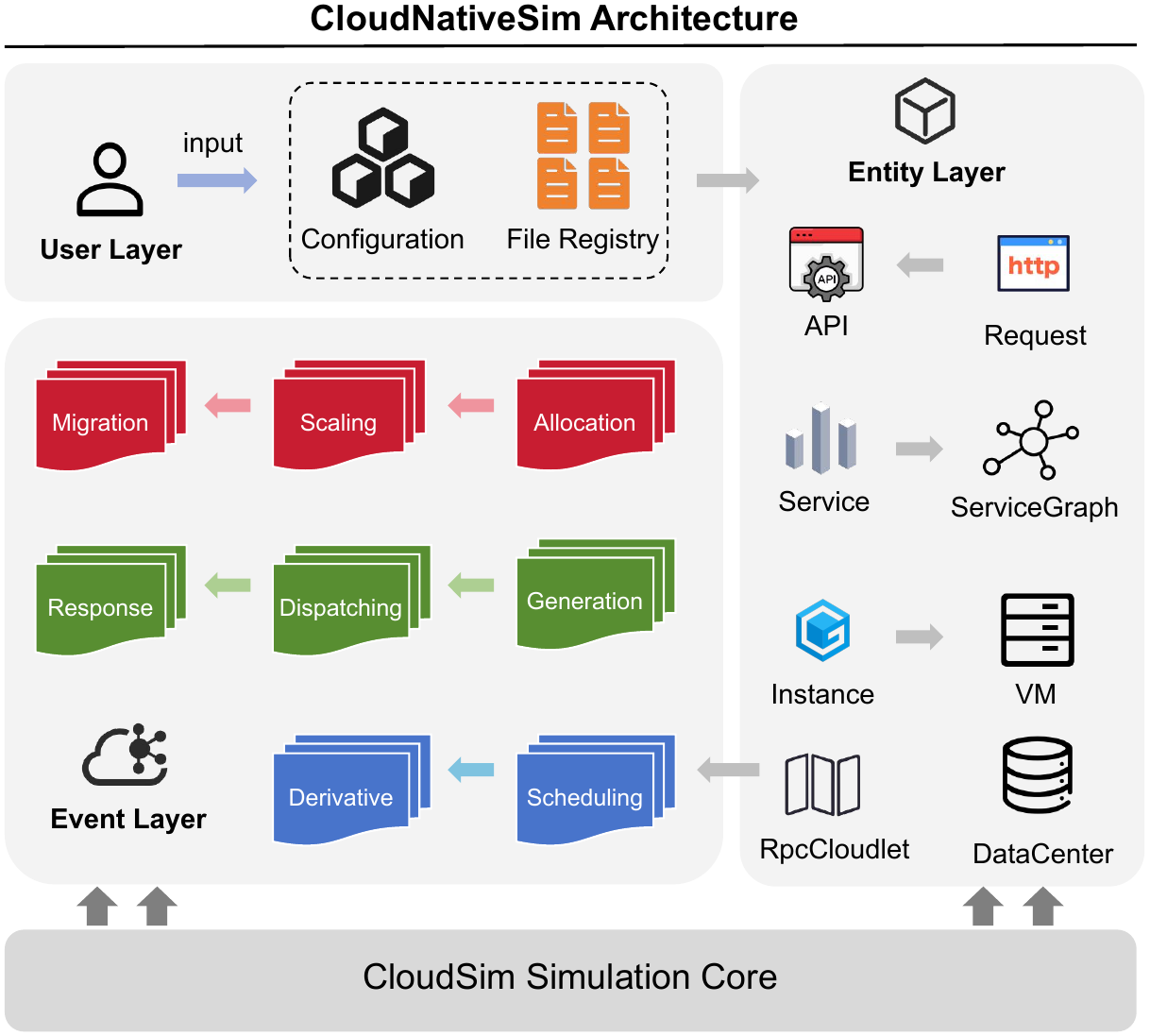} 
\caption{The architecture of CloudNativeSim.} 
\label{fig:framework}
\vspace{-1em}
\end{figure}

To ensure the simulator's flexibility and broaden the scope of modeling for potential use cases, CloudNativeSim has been developed using a discrete-event simulation approach. As illustrated in Figure \ref{fig:framework}, the architecture of CloudNativeSim is structured in layers: (i) The User Layer is dedicated to user interaction, emphasizing a streamlined interface to significantly enhance the user experience. (ii) The Event Layer is responsible for scheduling discrete events, enabling the simulator to handle a broader spectrum of scenarios efficiently. (iii) The Entity Layer consists of various entities representing the modeling of microservice-based architecture within the simulator; this layer also serves as the foundation for executing discrete events. (iv) The bottom layer integrates core components from the CloudSim toolkit \cite{cloudsim}, providing a robust and efficient simulation environment.

\subsection{User Layer: file registry and QoS feedback}\label{sec:user layer}

In the rapidly evolving field of cloud-native applications, optimizing user interaction is particularly important \cite{Hewage2023CloudSimEA}. At the user level, our primary task is to accurately understand user needs and the specific configurations required by the simulator. To enable users to tailor CloudNativeSim scenarios to their requirements, we offer flexible customization options and design a series of examples to facilitate testing and provide instructive demonstrations. The user layer provides options including deployment specifications, logging preferences, policy configurations and cloudlet parameters. Users can also tailor the reporting format to output specific metrics to either terminal displays or CSV files for subsequent analysis.

\begin{figure}[htb]
\centering
    \begin{subfigure}[b]{0.48\linewidth}
        \centering
        \includegraphics[width=0.75\linewidth]{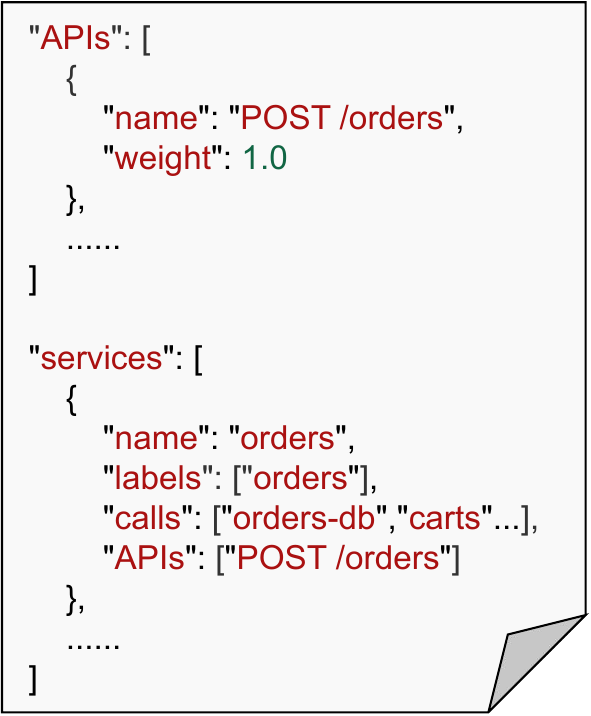} 
        \caption{JSON illustration of APIs and services configuration}
    \end{subfigure}
    \begin{subfigure}[b]{0.48\linewidth} 
        \centering
        \includegraphics[width=0.75\linewidth]{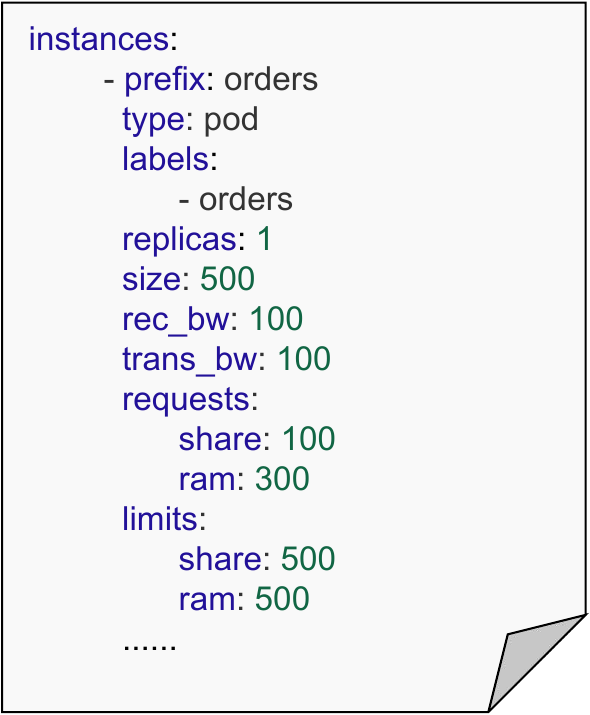} 
        \caption{YAML illustration of instance configuration}
    \end{subfigure}
\caption{A demonstration of documents used for file registry in CloudNativeSim.}
\label{fig:registry}
\vspace{-1em}
\end{figure}

Furthermore, in order to lower the barrier for users to describe configurations of cloud-native applications, CloudNativeSim employs a file-based method for entity registration. Figure \ref{fig:registry} illustrates a demo of the entity registry in CloudNativeSim, detailing configurations in both JSON and YAML formats. In the JSON section, APIs and services are defined. The API "POST /orders" has a weight of 1.0, while the "orders" service is labeled "orders" and calls services like "orders-db" and "carts". The YAML configuration specifies instance settings, including the "orders" pod, its labels, replicas, size, and bandwidth settings. It also details resource requests and limits for shares and RAM. These configurations provide a clear framework for managing APIs, services, and instances, ensuring efficient resource allocation and robust simulation of cloud-native environments. By focusing solely on the content of configuration files, users can operate the simulator without requiring an in-depth understanding of its mechanisms \cite{silva2014automation}. 

To more effectively manage and analyze the performance of cloud-native applications, CloudNativeSim provides dynamic QoS feedback through continuous monitoring and recording of performance metrics throughout the simulation runtime. This includes traditional instance-based metrics such as resource utilization, as well as detailed service-oriented metrics including interface-specific response times, Requests Per Second (RPS), and Service Level Objective (SLO) violation rates at both instance and system levels. All these metrics are displayed in real-time through terminal output, with configurable sampling frequencies to control reporting granularity. Users can further customize the reporting format by selecting specific metrics for display or exporting comprehensive logs to CSV files for detailed post-simulation analysis. These comprehensive monitoring capabilities enable thorough evaluation of cloud-native application behavior under various operational conditions, ensuring adherence to predefined service standards.

\subsection{Event Layer: the core design for scheduling}\label{sec:event layer}

Event layer is the core design of discrete event simulation, with the scheduling of events being centrally controlled by the application and data center. The detailed introduction are as below:

\begin{itemize}
    \item \textbf{Allocation:} After services and instances are registered, CloudNativeSim automatically initiates the allocation event. Allocation involves provisioning instances with appropriate resources and mapping services to instances and virtual machines, following a bottom-up process.
    \item \textbf{Scaling \& Migration:} As tasks are processed, the system status is dynamic. To maintain workload balance, CloudNativeSim needs to update the allocation mapping using scaling policies and migration mechanisms. The simulator includes three common scaling policies. These scaling policies typically involve migration mechanisms, especially when the scaling query cannot be completed with the current virtual machine.
    \item \textbf{Generation:} To simulate request arrivals from different clients in cloud-native applications, CloudNativeSim incorporates probability distributions (e.g. normal distribution) to generate cloudlets and requests. This approach requires users to configure only minimal parameters, enhancing the simulator's flexibility and enabling the efficient calculation of QoS metrics.
    \item \textbf{Dispatching:} Once a large volume of requests is generated, the simulator must have the capability to distribute these requests across different services. This process helps us understand the logic from request handling to response. Requests are characterized by their specific application interfaces (APIs), and are mapped to corresponding services. The simulator then selects suitable instances within each service based on load-balancing logic to distribute the requests.
    \item \textbf{Scheduling:} Cloudlet scheduling is foundational for the simulator's computations. CloudNativeSim employs the redesigned cloudlet scheduler module to ensure that response time and usage history calculations are efficient and dynamic.
    \item \textbf{Derivative:} When a service completes the cloudlets assigned to it, it automatically derives new cloudlets to next nodes along the service chain. This process ensure the number of cloudlets within the simulator to vary with the number of requests and services, and it also forms the basis for calculating response time using the critical path.
\end{itemize}

The Event Layer in CloudNativeSim is crucial for efficient resource management and modeling the dynamics of cloud-native applications. Events like allocation, scaling, and migration facilitate efficient resource management. Generation, dispatching, and response events manage the communication between clients and applications. To ensure the internal dynamics of applications, CloudNativeSim uses scheduling and derivation events to control the processing of cloudlets. This comprehensive approach enhances the accuracy of QoS metrics calculation, improves flexibility and usability, and ensures compatibility with CloudSim users. Ultimately, it boosts the efficiency and scalability of the simulator.

\subsection{Entity Layer: data structure and mapping logic}\label{sec:entity layer}

The design of the Entity Layer in CloudNativeSim efficiently models the entities of cloud-native applications. Requests are generated by clients to access specific APIs and serve as the sources of tasks that propagate through the service chain. Services and instances are the most crucial data structures in a microservice-based architecture, responsible for the virtualization of functional modules. Services are a logical concept that map to a set of instances with the same functionality, and most scheduling events in CloudNativeSim operate based on services. Furthermore, services form mutual dependencies, creating service chains and service graphs. CloudNativeSim also extends the RpcCloudlet data structure to represent communication between services and to handle request processing. Additionally, CloudNativeSim integrates core entities from CloudSim 3, including VMs and DataCenters, with necessary extensions to enhance their functionality. More modeling details will be discussed in Section \ref{sec 4}.

The mapping mechanism is fundamental for associating and updating these entities. In CloudNativeSim, this mechanism is implemented through a HashMap, managing relationships between different entities based on names or labels. For example, it maps requests to APIs and services to the service graph. Maintaining these mapping relationships reduces redundancy in update operations and enhances the efficiency and flexibility of the simulator. 

To further enhance the usability and compatibility of the simulator, CloudNativeSim adopts the resource profiling approach of CloudSim, enabling seamless migration between the new and old simulators. In terms of resource profiling, various service instances, such as pods and containers, are assigned a range of resource attributes, including ID, Million Instructions Per Second (MIPS), memory (RAM), and bandwidth (BW). This approach not only preserves the familiarity for original CloudSim users but also ensures the scalability and extensibility of the simulator.

By incorporating these design elements, CloudNativeSim provides a robust and flexible environment for simulating cloud-native applications. This ensures efficient resource management, accurate modeling of service interactions, and seamless integration with existing simulation frameworks, ultimately enhancing the overall effectiveness and adaptability of the simulator.

\subsection{Implementation of Simulation Components}\label{sec:implementation}

\begin{figure}[htb] 
\centering 
\includegraphics[width=0.85\textwidth]{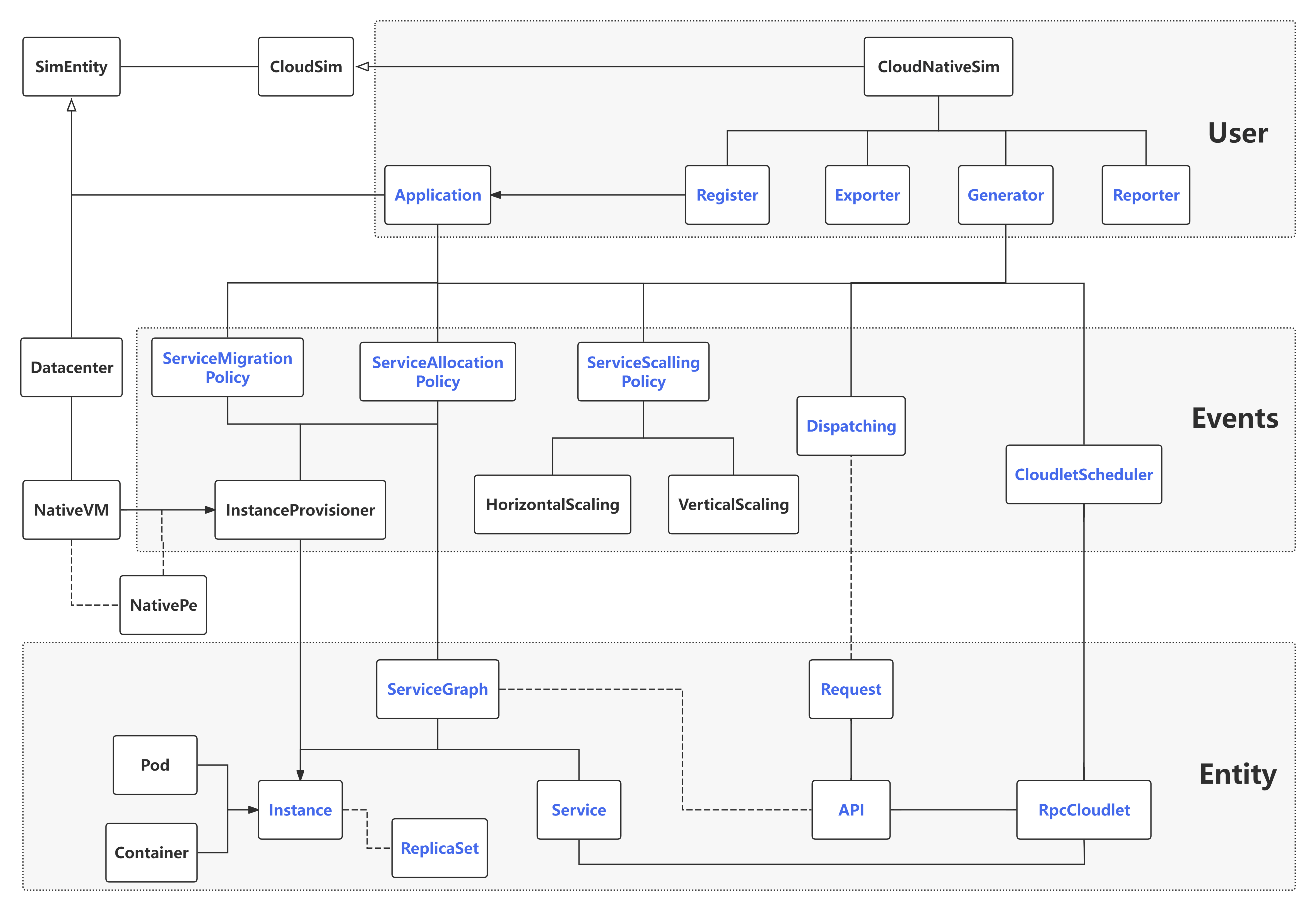} 
\caption{The class diagram of CloudNativeSim detailing levels of user interaction, event management, and entity representation in a cloud simulation environment.} 
\label{fig:class}
\vspace{-1em}
\end{figure}

The main class diagram for CloudNativeSim is illustrated in Figure \ref{fig:class}. This diagram is organized vertically, corresponding sequentially to the three hierarchical levels depicted in Figure \ref{fig:framework}. At the event level, specific events initiate the scheduling of SimEvent operations. CloudNativeSim has been developed using Java 17, inheriting the robust discrete event simulation framework from the core of CloudSim.

Below are the detailed introduction of the main classes (highlighted in blue in Fig. \ref{fig:class})  and their functionalities within the CloudNativeSim architecture:

\begin{enumerate}
\item \textbf{Application}: This class serves as the cornerstone of CloudNativeSim, encapsulating a cloud-native application. It extends from the SimEntity class and is activated post the completion of resource configuration at the VM layer. Principal responsibilities include linking entities with data centers and orchestrating the scheduling and collaboration of events.

\item \textbf{Register and Reporter}: These classes aim to enhance the system's input and output capabilities. They streamline user interactions and are crucial in generating and disseminating QoS metrics. The \textit{Register} class handles the registration of various entities, while the \textit{Reporter} class collects and reports performance metrics, ensuring users have comprehensive insights into the system's behavior. Details in Section \ref{sec:user layer}.

\item \textbf{Exporter}: This class is responsible for collecting various metrics during the simulator's operation and calculating QoS-related outputs based on these metrics. The \textit{Exporter} ensures that performance data is accurately captured and made available for analysis, contributing to the system's transparency and accountability.

\item \textbf{Service and ServiceGraph}: The \textit{Service} class represents the concept of a service. Each \textit{Service} object is a node in the \textit{ServiceGraph}, which is a DAG data structure. The \textit{ServiceGraph} maps out the relationships and dependencies between various services, facilitating efficient service orchestration and interaction. Details in Section \ref{sec:DAG}.

\item \textbf{RpcCloudlet and CloudletScheduler}: \textit{RpcCloudlet} is an extension in CloudNativeSim that represents inter-service communication tasks. It defines attributes for computing latency and resource usage. The \textit{CloudletScheduler} orchestrates cloudlets within the current service, distributing them to different instances for processing based on the load balancing status of the instances. Details in Section \ref{sec:cloudlets} and Section \ref{sec:scheduler}.

\item \textbf{Generator and Dispatching}: The \textit{Generator} class dynamically generates requests and cloudlets on behalf of clients. Once requests reach the API, they are dispatched to various service chains, while cloudlets are managed by the \textit{CloudletScheduler}. This mechanism ensures that client requests are efficiently processed and directed to the appropriate services. Details in Section \ref{sec:generator}.

\item \textbf{Request and API}: These classes simulate user interactions with cloud-native applications. The \textit{Request} class represents user access to the application, while the \textit{API} serves as the entry point for these requests. The \textit{API} maps requests to different service chains, and as requests enter the system, they generate cloudlets that are processed along these chains. 

\item \textbf{Instance and ReplicaSet}: These classes contain the necessary fields for instances. CloudNativeSim internally implements Pod and Container instances, allowing for user customization. A \textit{ReplicaSet} is a collection of instances with identical properties, used to achieve horizontal scaling. This setup ensures that multiple instances can be managed collectively to handle increased loads.

\item \textbf{ServiceAllocationPolicy and ServiceMigrationPolicy}: These classes abstract the algorithmic frameworks necessary for the deployment and migration of services. They work in tandem with the \textit{InstanceProvisioner} class to facilitate the efficient deployment of services and instances within the simulation environment. \textit{ServiceAllocationPolicy} focuses on initial service placement, while \textit{ServiceMigrationPolicy} manages the dynamic relocation of services to optimize resource utilization. Details in Section \ref{sec:allocation}.

\item \textbf{ServiceScalingPolicy}: \textit{ServiceScalingPolicy} is the core module for service scheduling, this class is equipped with algorithms designed for both horizontal and vertical scaling. It enables dynamic service management based on operational demands, allowing the system to adapt to varying loads by scaling resources up or down as needed. Details in Section \ref{sec:scaling}.

\end{enumerate}

To ensure the extensibility of the simulator, CloudNativeSim inherits from CloudSim 3.0, which allows it to integrate seamlessly with classical component modeling. The strategic decision to use CloudSim 3.0 as the base instead of higher versions is due to the fact that the latest versions of CloudSim, although powerful, introduce many complex features. These features often complicate the framework and decrease its extensibility. By using the simpler, more robust framework of CloudSim 3.0, it becomes easier to extend and customize. Thus, CloudNativeSim is able to simulate a broader range of microservice-based architectures for cloud-native applications.

\section{Modeling of Cloud-native Applications}\label{sec 4}


To comprehensively simulate the performance of a cloud-native application, it is necessary to model its internal structure. During the modeling process, the simulator must address the challenges posed by the heterogeneity and high dynamism inherent in microservice-based architecture. Heterogeneity refers to services operating in different code environments and are mapped to various instances, forming a complex entity communication network. Additionally, high dynamism, driven by varying request workloads, complicates the prediction of resource and state changes.

\begin{table}[htbp]
    \centering
    \renewcommand{\arraystretch}{1.25} 
        \begin{tabular}{|c|m{6cm}||c|m{5.25cm}|} 
        \hline
        \textbf{Terms} & \textbf{Definitions} & \textbf{Terms} & \textbf{Definitions} \\
        \hline
        $service$ & A service is an independent unit that performs functionality and communicates with other services via APIs. &
        $instance$ & An instance refers to a specific performer of a service, such as a pod or container. \\
        \hline
        $task$ & A task corresponds to a request that invokes multiple services and is completed by several cloudlets working together. The task length is the total computational workload measured in millions of instructions (MI). &
        $cloudlet$ & A cloudlet is a fundamental computational unit used for task processing in the simulation. The cloudlet length is also measured in MI. \\
        \hline
        $instanceList$ & A collection of all active service instances in the system at a given time. &
        $scalingList$ & A dynamic list of instances marked for scaling operations (scale-out or scale-in). \\
        \hline
        $apiList$ & A list of all available APIs exposed by services in the system. &
        $\{w\}$ & A set of weights representing the probability distribution for API selection by clients. \\
        \hline
        $P$ & The set of all possible execution paths through the service chain, where each path represents a sequence of service invocations. &
        $CP$ & The critical path represents the longest execution sequence in terms of response time among all possible paths $P$. \\
        \hline
        $c_i$ & The $i$-th client entity in the system that generates service requests according to defined patterns. &
        $N_c$ & The current total number of active clients in the system at time $t$. \\
        \hline
        $v$ & The client growth rate, defining how $N_c$ changes over time ($v > 0$ for growth). & 
        $p$ & The inter-request interval range $p = [p_0, p_1]$, where $p_0$ and $p_1$ represent the minimum and maximum waiting time between consecutive requests. \\
        \hline
        $timeLimit$ & The maximum time period during which clients are allowed to send requests. &
        $numLimit$ & The maximum number of requests to be processed in the simulation. \\
        \hline
        $\lambda$ & The request arrival rate (QPS), measuring the number of incoming requests per second. &
        $R$ & The cumulative total number of requests since the start of simulation until the current time point. \\
        \hline
        \end{tabular}
        \caption{Terms and definitions for cloud-native application modeling in CloudNativeSim.}
        \label{tab:terms}
    \end{table}

To facilitate clear communication and understanding, we first introduce the key terms and their definitions used throughout this paper in Table \ref{tab:terms}. These terms form the foundation of our modeling approach and will be referenced extensively in subsequent sections.

In response to the above challenges, CloudNativeSim employs a novel modeling approach, which consists of three key components: (i) The service graph modeling constructs dependencies between services, effectively describing the communication topology among heterogeneous entities. (ii) The cloudlet scheduler provides multi-level queues to manage cloudlets, ensuring flexible resource and status updates within the system. (iii) The client request processing module simulates dynamic request generation and calculates response time by updating the critical path, capturing the dynamic interactions between cloud-native applications and clients.

\subsection{Modeling of Service Graph}

\begin{figure}[htbp]
    \centering
    \includegraphics[width=0.95\textwidth]{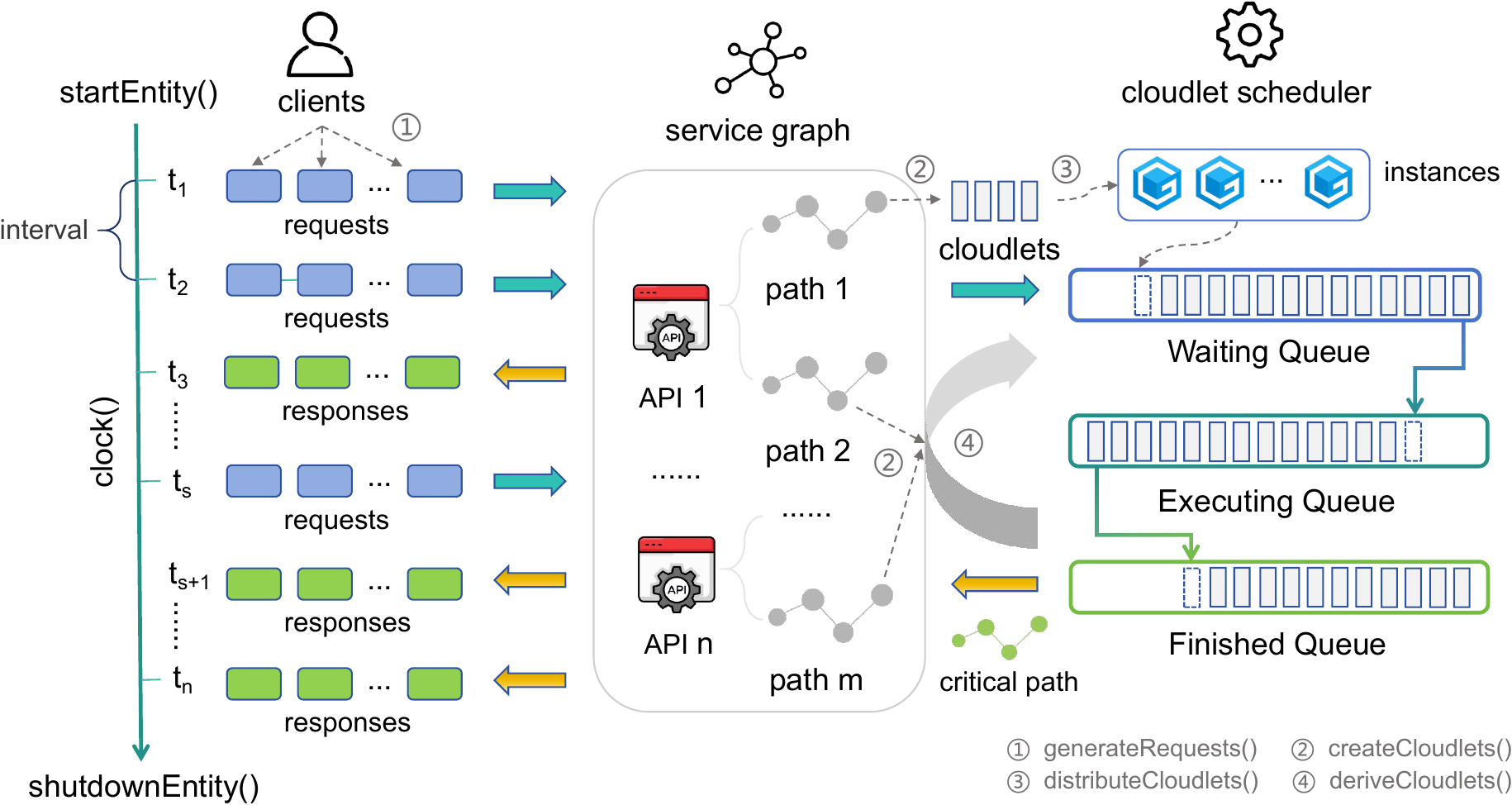}
    \caption{Modeling approach divided into three components: client requests modeling, service graph modeling, and cloudlet scheduler. Numbers correspond to built-in methods.}
    \label{fig:modeling}
    \vspace{-1em}
\end{figure}

To accurately represent communications between microservice entities, considering that different entities are mapped to various services with specific dependencies and communication processes, it is essential to model the service graph comprehensively. As shown in the second part of Figure \ref{fig:modeling}, when requests arrive the APIs, they will call different chains that compose a service graph and create cloudlets for entity communications. CloudNativeSim utilizes a DAG to depict service dependencies efficiently and employs remote procedure calls (RPC) to simulate communication processes. By integrating these advanced modeling techniques, CloudNativeSim can significantly enhance both the accuracy and efficiency of service communication simulations.

\subsubsection{Service Dependency Modeling based on DAG} \label{sec:DAG}

Given that communications between service communication is unidirectional and lacks cyclic calls, using a DAG structure is an efficient approach for modeling service dependencies \cite{firm,crisp,Ma2018DAG}. This structure enables more accurate and detailed simulations of service interactions and dependencies, providing valuable insights into system performance and potential failure points. By utilizing DAG, CloudNativeSim can help developers and researchers better understand the complex dynamics of cloud-native applications, which leads to optimized design and enhanced strategies.

\begin{figure}[htbp]
    \centering
    \includegraphics[width=0.8\linewidth]{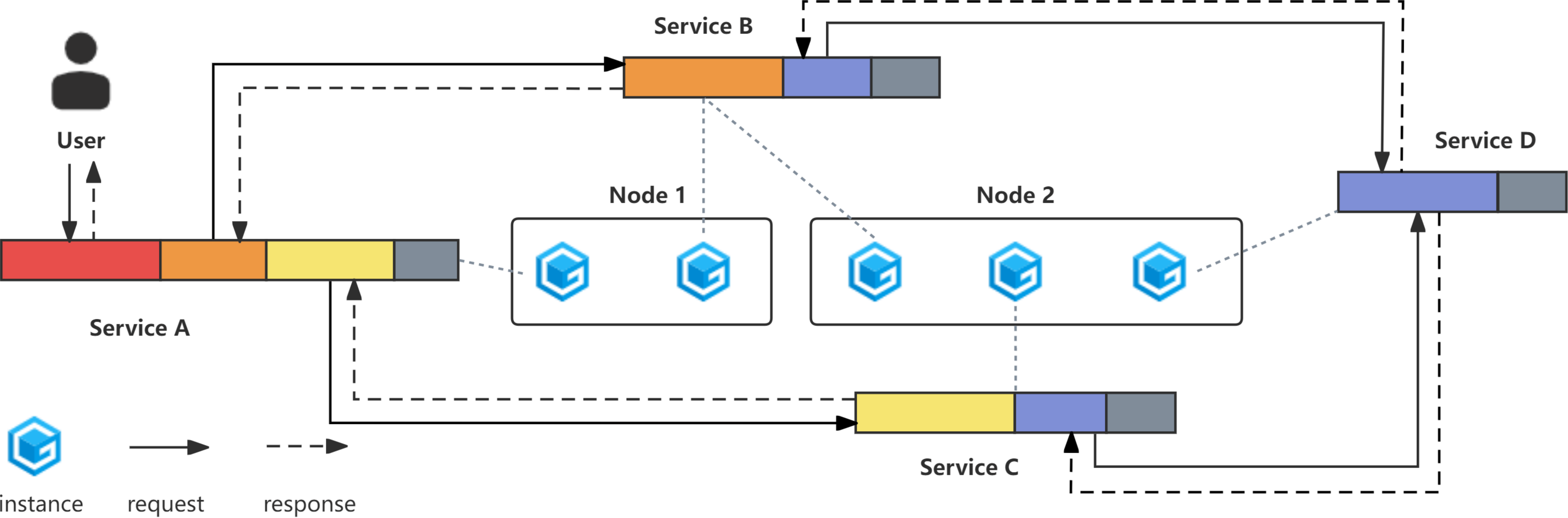}
    \caption{A simple illustration of service dependencies using a DAG structure, depicting the flow of requests and responses. Additionally, it shows how services are mapped to instances, which are distributed across different nodes.}
    \label{fig:DAG}
    \vspace{-1em}
\end{figure}

\begin{figure}[htbp]
    \centering
    \includegraphics[width=0.7\linewidth]{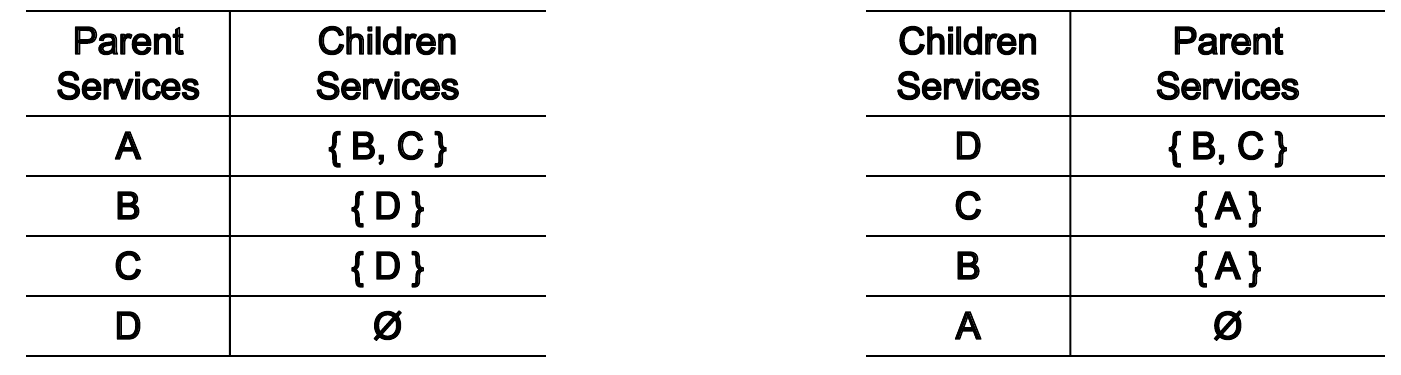}
    \caption{The data structure of service dependency tracking in CloudNativeSim. The left table implements forward lookup where each service maps to its dependent services. The right table implements reverse lookup where each service maps to services that call it.}
    \label{fig:reverse}
    \vspace{-1em}
\end{figure}

Figure \ref{fig:DAG} presents a simple DAG to depict the service dependencies: Service A calls Services B and C, which in turn call Service D. Each service comprises many code segments represented by rectangles of different colors. The DAG structure includes multiple service chains corresponding to APIs, and client requests will propagate along these chains, forming the inter-service communications and eventually responding to the clients. These chains can be recursively defined according to different source services. Figure \ref{fig:reverse} shows the built-in data structures of CloudNativeSim, representing service dependencies through a DAG implementation. By utilizing these reverse-linked lists, our simulator can efficiently track service interactions and trace call chains in both forward and reverse directions.

On the other hand, Figure \ref{fig:DAG} also demonstrates the vertical mapping relationship between services and instances. When a request arrives, services use the mapping mechanism mentioned in Section \ref{sec:entity layer} to allocate the corresponding tasks to multiple instances. Each instance is deployed on different nodes, such as virtual machines and physical machines. Essentially, the response process of requests involves utilizing node resources to execute tasks.

By modeling the service graph through a DAG structure, CloudNativeSim offers a clear and detailed representation of service dependencies. This approach enhances user insights into cloud-native applications, making internal changes more transparent and easier to understand.

\subsubsection{Inter-service Communication Modeling based on RpcCloudlet}\label{sec:cloudlets}


Inter-service communication efficiency serves as a critical determinant in modern distributed systems' performance \cite{Weerasinghe2022EvaluatingTI}. The emergence of gRPC \footnote{https://grpc.io/} has revolutionized this domain by introducing a high-performance, language-agnostic communication protocol \cite{gRPC}. Drawing inspiration from these advances, CloudNativeSim introduces RpcCloudlet as a novel modeling construct to simulate inter-service communications. This extension builds upon CloudSim's fundamental Cloudlet abstraction while incorporating contemporary distributed computing paradigms.

\begin{figure}[htbp]
\centering
\begin{subfigure}[b]{0.48\linewidth} 
\includegraphics[width=\linewidth]{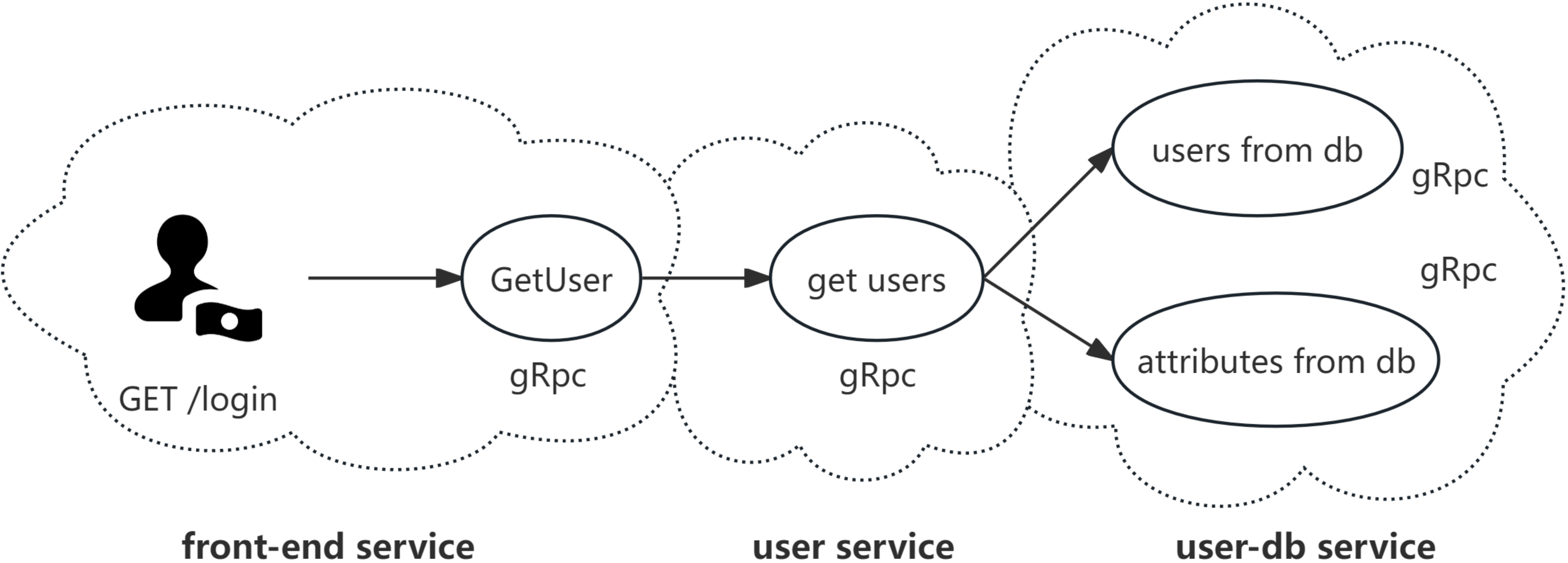}
\caption{Request invocation sequence in a physical service environment}
\label{fig:endpoints-1}
\end{subfigure}
\hfill
\begin{subfigure}[b]{0.48\linewidth} 
\includegraphics[width=\linewidth]{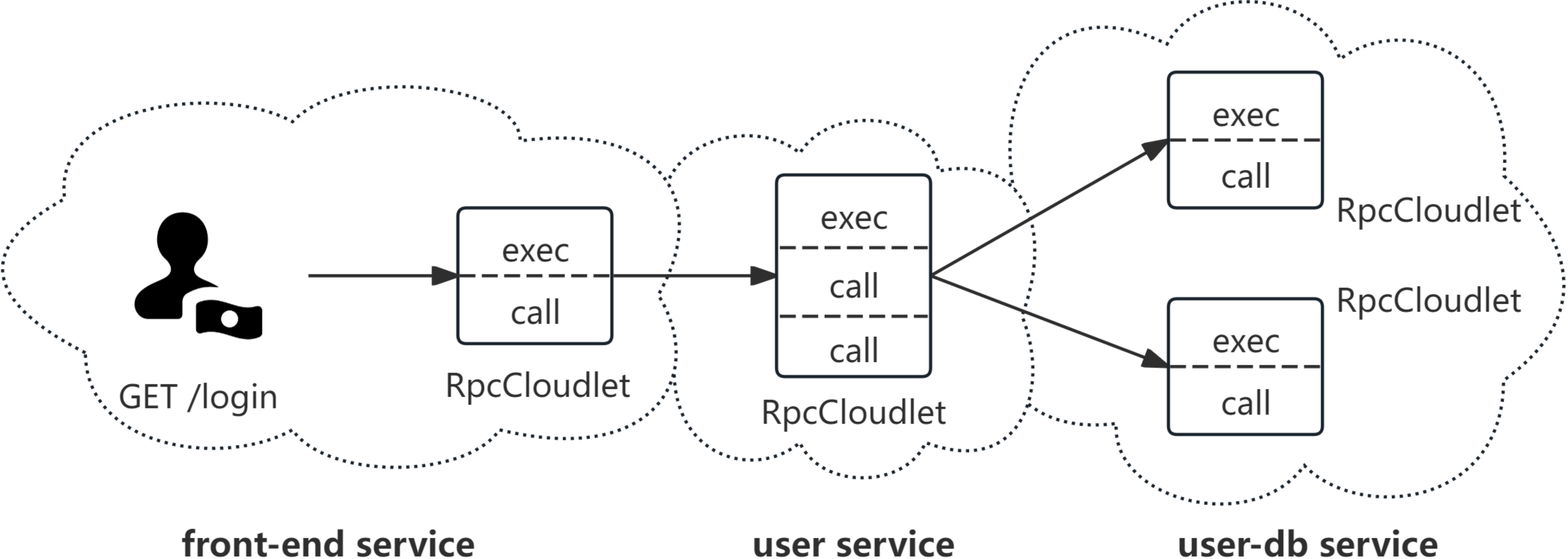}
\caption{Request invocation sequence in a simulated service environment}
\label{fig:endpoints-2}
\end{subfigure}
\caption{Comparison of request invocation sequence in physical and simulated environments.}
\vspace{-1em}
\end{figure}

The architectural distinction between physical systems and their simulated counterparts becomes evident when examining their respective communication patterns. Figure \ref{fig:endpoints-1} illustrates this through Jaeger's \footnote{https://www.jaegertracing.io/} distributed tracing visualization of the sockshop microservices application. In production environments, a "\textit{Get /login}" API request initiates a cascade of service invocations, where each endpoint (represented as ovals) belongs to distinct microservices, forming intricate service dependency chains. Conversely, Figure \ref{fig:endpoints-2} demonstrates the simulator's approach, where inter-service communications are abstracted through dynamically generated cloudlets, offering a more granular representation of these complex interactions.

Our implementation marks a significant difference from CloudSim's traditional static cloudlet allocation model. Where CloudSim employs upfront cloudlet definition and assignment to virtual machines—an approach that fails to capture the dynamic nature of modern microservices—our system introduces an innovative "dynamic derivation" mechanism. This approach generates cloudlets on-demand, precisely when their corresponding services are invoked, establishing a more realistic representation of service interactions. A distinguishing feature of this design is its ability to accommodate multiple cloudlet instances for a single service throughout the request lifecycle. As illustrated in Figure \ref{fig:endpoints-2}, the user-db service exemplifies this capability, spawning multiple cloudlet instances in response to various service invocations. This dynamic instantiation pattern closely aligns with the principles of gRPC communication, hence the designation "RPCCloudlet."

To address the inherent complexity of request processing, our system implements a novel cloudlet management mechanism. Through empirical observation, we discovered that computational requirements exhibit statistical independence, making them amenable to modeling via Gaussian distribution. The cloudlet configuration framework incorporates user-definable parameters, including expected mean length, variance, and service-instance mapping specifications. Upon parameter initialization, the system autonomously derives new cloudlets as service invocations occur.

This comprehensive approach to inter-service communication modeling offers several compelling advantages: First, the cloudlet derivation process inherently reflects service dependencies, enhancing simulation transparency and interpretability. Second, the integration with a discrete event framework enables dynamic performance metric evolution, particularly in latency measurements, contributing to enhanced system adaptability. Third, the design preserves fundamental RPC characteristics, ensuring that individual cloudlet failures do not compromise overall system execution, thereby maintaining robust operational continuity and system resilience. These features collectively establish a more realistic and sophisticated simulation environment for modern distributed systems.


\subsection{Design of Cloudlet Scheduler} \label{sec:scheduler}

To ensure high dynamism in inter-service communication during simulation, we redesign the cloudlet scheduler module, as depicted in the third part of Figure \ref{fig:modeling}. Given that services are the primary entities in the communication network and that cloudlets are transmitted between these services, we implement a scheduler for each service to manage its cloudlets. Each cloudlet scheduler includes a waiting queue, execution queue, and finished queue to manage the flow of cloudlets. The specific functionalities are as follows:

\begin{itemize}
    \item \textbf{Waiting Queue}: When a new cloudlet is created by a service, it is initially added to the waiting queue, awaiting idle resources. The cloudlets in this queue are sorted based on predefined principles, such as first-come-first-served or priority-based service.
    \item \textbf{Execution Queue}: If the scheduler detects that the current service has sufficient resources, it will allocate the waiting cloudlets to appropriate instances based on load balancing decisions and move them to the execution queue. Cloudlets not selected for execution will re-enter the waiting queue, with their waiting time increased.
    \item \textbf{Finished Queue}: Upon completion, cloudlets are moved to the finished queue, where their execution results and status information are recorded. This data serves as the basis for subsequent analysis and optimization of scheduling strategies.
\end{itemize}

To prevent instances from becoming imbalanced and congested, load balancing decisions must be made when cloudlets transition from the waiting queue to the execution queue. CloudNativeSim includes built-in methods for load balancing, such as selecting the instance with the maximum idle resources or using a random allocation method. Both approaches efficiently balance the load across instances. Additionally, users of the simulator can implement custom load balancing strategies via the provided interfaces.

During execution, cloudlets use time-sharing CPU resources among instances as default, where each instance alternately executes multiple cloudlets in time slices. Users may also choose more complex methods, such as spatial multiplexing. In CloudNativeSim, we provide two default choices: equal time slice multiplexing and unequal time slice multiplexing, both affecting the instance selection process. For equal time slices, methods such as round-robin or random can be used. For unequal time slices, methods like best-effort are available. Users can customize selection strategies based on system load, the current state of instances, and cloudlet priority.

This approach enables CloudNativeSim to accurately simulate service request processing while providing flexible and efficient resource management. By detecting resource-starved cloudlets and dynamically recreating and scheduling new cloudlets, CloudNativeSim helps users optimize and manage their systems more efficiently.

\subsection{Processing of Client Requests}

In the simulation of cloud-native applications, user requests serve as the core entry point, driving all scheduling activities. Consequently, processing these requests becomes the primary objective of the simulation. Compared to traditional solutions, this request-oriented simulation emphasizes updating and providing feedback on request-related metrics (including latency, RPS, and SLO violation rates). To achieve this, we need to describe two critical components in CloudNativeSim: request generation and response time calculation.

\subsubsection{Request Generation Modeling}\label{sec:generator}
To dynamically simulate the arrival of requests, CloudNativeSim employs a more versatile approach in which the arrival rate is dynamically dictated by the system's parameters. The Request Generator component simulates user request arrivals based on specified parameters, such as the number of clients ($N_c$), client growth rate ($v$), waiting time intervals ($p$), request weight for each API ($\{w\}$), list of APIs ($apiList$), maximum generation time ($timeLimit$), and maximum number of requests ($numLimit$). The client growth rate ($v$) represents the rate at which the number of clients increases over time, acting as a velocity factor that dictates how quickly new clients are added to the system. Waiting time intervals ($p$) define the range of time intervals that clients wait between sending consecutive requests, represented as an interval $p = [p_0, p_1]$, where $p_0$ is the minimum wait time and $p_1$ is the maximum wait time. Each client will randomly select a wait time from this interval after sending a request. Request weight for each API ($\{w\}$) is a set of weights that represent the probability distribution for selecting different APIs when generating requests. Each weight corresponds to an API in the $apiList$, determining how likely it is for a particular API to be chosen when a client sends a request.

\begin{algorithm}
\caption{Request Generation Algorithm}
\label{al:generator}
\hspace*{0.02in} {\bf Input:}
$N_c$ (number of clients), $v$ (client growth rate), $p$ (waiting time intervals), $\{w\}$ (weight set), $apiList$ (list of APIs), $timeLimit$ (maximum generation time) and $numLimit$ (maximum number of requests) \\
\hspace*{0.02in} {\bf Output:}
$new\_requests$ (new generated requests)

\begin{algorithmic}[1]
\State Initialize $waitingClients$ as a list of size $N_c$, all elements set to 0;
\State Initialize $currentTime$ and $currentNum$ to 0;

\While{$currentTime < timeLimit$ \textbf{and} $currentNum < numLimit$}
    \If{$currentClients < N_c$}
        \State Set $currentClients \gets \min(v \times currentTime, N_c)$;
    \EndIf

    \For{each $i$ from $0$ \textbf{to} $currentClients - 1$}
        \If{$waitingClients[i] == 0$}
            \State Set $api \gets $ randomly select an API from $apiList$ according to $\{w\}$;
            \State Set $new\_requests \gets \text{createRequests}(api, currentTime)$;
            \State Send $new\_requests$ to the system;
            \State Increment $currentNum$ by the number of $new\_requests$;
            \State $waitingClients[i] \gets$ randomly select a wait time from interval $p$;
        \Else
            \State Decrement $waitingClients[i]$ by 1;
        \EndIf
    \EndFor

    \State $currentTime$ ++;
\EndWhile
\end{algorithmic}
\end{algorithm}

When designing the Request Generator component, we were inspired by Locust \footnote{https://locust.io/} and aimed to simplify it for simulation purposes. Algorithm \ref{al:generator} describes the logic of this component: at the beginning of the simulation, users need to specify several parameters: the final number of clients $N_c$, client growth rate $v$, request weight for each interface $w$ (default is 1), user request waiting span $p$, time limit $timeLimit$, or quantity limit $numLimit$. The request generation process is divided into the following steps:

\begin{enumerate}
\item Users configure and initialize the API interface list and other parameters (lines 1-2).
\item The number of clients increases from 0 to $N_c$ at a rate of $v$ (lines 4-6).
\item As the number of clients grows, each client randomly selects an interface to send requests according to the assigned weight (lines 9-12).
\item After sending requests, clients configure a waiting period, with the duration randomly selected within the span $p$ (line 13).
\end{enumerate}

Based on the above logic, we can model some key metrics mathematically. The first is the number of clients $N(t)$, which initially grows linearly with time $t$ and remains constant after reaching $N_c$:

\begin{align}
\label{eqa:N(t)}
N(t) = \min(N_c, v \cdot t).
\end{align}

Next, the request arrival rate $\lambda(t)$ refers to the number of requests arriving per second. Considering that each client waits for a random period after sending a request, the distribution is no longer a simple Poisson process. However, when the number of requests is sufficiently large, this waiting time is uniformly distributed within the interval $p = [p_0, p_1]$. Therefore, we can use the expected value of the waiting time $E(p)$ to approximate the request arrival density:

\begin{align}
E(p) = \frac{p_0 + p_1}{2}.
\end{align}

Thus, the request arrival rate $\lambda(t)$ at time $t$ is:

\begin{align}
\label{eqa:qps}
\lambda(t) = \frac{N(t)}{E(p)} = \min(N_c, v \cdot t) \times \frac{2}{p_0 + p_1}.
\end{align}

Finally, the total number of requests $R(t)$ represents the cumulative number of requests generated from the start of the simulation up to a certain time point. To find $R(t)$, we integrate the request arrival rate $\lambda(t)$. Since $\lambda(t)$ has different expressions at different stages, piecewise integration is required:

\begin{equation}
\label{eqa:R(t)}
R(t) = 
\begin{cases} 
\frac{v }{\left(p_0 + p_1\right)} \cdot t^2 & \text{if } t \leq \frac{N_c}{v} \\[10pt]
\frac{2N_c}{\left(p_0 + p_1\right)} \cdot t - \frac{N_c^2}{v  \left(p_0 + p_1\right)} & \text{if } t > \frac{N_c}{v}
\end{cases}.
\vspace{0.25cm}
\end{equation}

By analyzing the above formulas, we can infer certain properties of these metrics, such as linearity or non-linearity, specific convergence values, and more. We will validate these properties in Section \ref{eva:generator}.

\subsubsection{Response Time Calculation and Critical Path Identification}\label{sec:response}


In cloud-native applications, the response time of a request is determined by the completion time of all its dependent tasks. CloudNativeSim employs a discrete event framework to track the execution of these tasks and calculate the response time. After obtaining the response time, the system identifies the critical path by backtracking through the service invocation chain, which helps users understand the performance bottlenecks and optimize resource allocation.

\begin{figure}[htbp]
    \centering
    \includegraphics[width=0.7\linewidth]{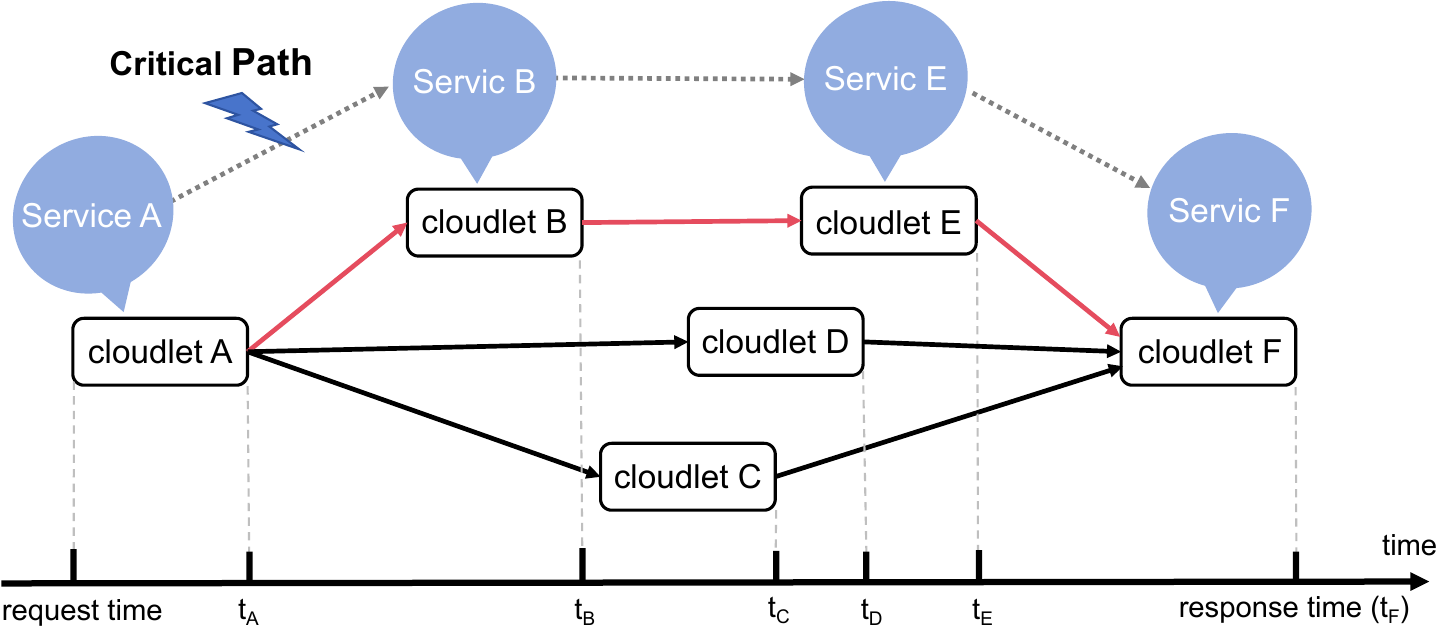}
    \caption{A demonstration for response time calculation and critical path identification. The red path (A→B→E→F) shows the identified critical path in service graph.}
    \label{fig:al2}
    \vspace{-1em}
\end{figure}

Figure \ref{fig:al2} illustrates this process with a service invocation graph. The nodes (A-F) represent individual cloudlets, with edges showing their invocation relationships. The discrete event framework tracks the execution of each cloudlet, recording their start and completion time. The response time is determined when the last cloudlet (F in this case) completes its execution. Subsequently, given the one-to-one correspondence between cloudlets and services, the system performs a post-hoc analysis by backtracking from F through the cloudlets with the latest completion time, thereby identifying the critical path (highlighted in red as A→B→E→F) in the service graph. This critical path represents the sequence of services whose execution time directly impact the overall response time, which is particularly valuable for users in analyzing bottlenecks in cloud-native applications.


\begin{algorithm}
\caption{Response Time Calculation and Critical Path Identification}
\label{al:cp}
\hspace*{0.02in} {\bf Input:} $request$ (the incoming request) \\
\hspace*{0.02in} {\bf Output:} $responseTime$ (the maximum of response time) and $CP$ (the critical path in service chain)

\begin{algorithmic}[1]
\State Initialize $responseTime$ to 0;
\State Initialize $serviceChain$ by $request.API$;
\For{each $path$ \textbf{in} $serviceChain$}
    \State Initialize $current\_delay$ to 0;

    \For{each node \textbf{in} path}
        \State Set $current\_delay \gets current\_delay + node.delay$;
    \EndFor
    \If{$current\_delay > responseTime$}
        \State Update $responseTime$ to $current\_delay$;
        \State Update $CP$ to $path$;
    \EndIf
\EndFor

\State Update the $serviceChain$ with $CP$;
\State \Return $responseTime$;

\end{algorithmic}
\end{algorithm}

Algorithm \ref{al:cp} demonstrates the critical path identification process. The algorithm examines all possible paths in the service chain to identify the one with the maximum accumulated delay. Let \( P \) denote the set of all paths in the service chain, and \( N_p \) denote the set of nodes in path \( p \in P \). For each path, the total delay \( D_p \) is calculated as:
\begin{align}
D_p &= \sum_{n \in N_p} \text{delay}(n),
\end{align}
where \( \text{delay}(n) \) represents the actual execution time of node \( n \) obtained from the discrete event simulation.

The critical path \( CP \) is then determined as the path with the maximum total delay:
\begin{align}
CP &= \arg\max_{p \in P} D_p.
\end{align}

This approach enables CloudNativeSim to provide valuable insights into service performance. By analyzing the critical path, users can identify which service invocations contribute most significantly to the overall response time. This information is particularly valuable for system optimization, as it highlights potential bottlenecks and suggests where performance improvements would be most effective.

\section{Implementation of Scheduling Policies}\label{sec 5}

In addition to the aforementioned challenges, a simulator with high extensibility also requires the implementation of reliable interfaces. In this section, we introduce the main scheduling policies that CloudNativeSim provides for cloud-native applications. These policies ensure extensibility while maintaining the dynamism of the simulation process.

\subsection{Service Placement Policy}\label{sec:allocation}

To ensure the system can adequately handle the requests received by applications, CloudNativeSim implements the interfaces of placement policies, which include the provisioner component, service allocation, and instance migration. An efficient implementation of these placement policies can provide a more accurate estimation of system resource utilization, which is highly beneficial for simulating cloud-native applications with a microservice-based architecture.

The provisioner component, which manages and tracks resource allocation and utilization, serves as the foundation for other policies in CloudNativeSim, providing the mapping of various resources such as CPU cycles, memory, and storage. During the simulation runtime, the provisioner component continuously calculates the total and available resources in real-time, as well as which entities are utilizing these resources.

CloudNativeSim adopts a bottom-up approach for implementing service allocation policy. Algorithm \ref{al:allocation} describes the default service allocation process in CloudNativeSim:

\begin{algorithm}
\caption{Service Allocation Algorithm}
\label{al:allocation}
\hspace*{0.02in} {\bf Input:}
$service$ (the service to be allocated), $vmList$ (list of available virtual machines), $instanceList$ (list of service instances) \\
\hspace*{0.02in} {\bf Output:} 
$result$ (the result of allocation)

\begin{algorithmic}[1]
\State Initialize $result \gets$ false;
\State Match instances in $instanceList$ to $service$ through their labels;
\State Initialize a $sortedQueue$ with $vmList$ by their available PE resources in descending order;
\For{each $instance$ \textbf{in} $instanceList$}
    \For{each $vm$ \textbf{in} $sortedQueue$}
        \If{$instance.require\_resources \leq vm.idle\_resources$}
            \State Allocate the $instance$ to $vm$;
            \State Set $instance.isAllocated \gets$ true;
            \State $result \gets$ true;
        \EndIf
        \State Update $vm.idel\_resources$;
    \EndFor
    \State Update $sortedQueue$;
    \State Update the provisioner in $instance$;
\EndFor
\State \Return $result$

\end{algorithmic}

\end{algorithm}

The algorithm begins by instantiating services based on predefined labels, creating a many-to-many mapping (line 2). Subsequently, CloudNativeSim manages virtual machines into a prioritized queue based on available processing element (PE) resources, ensuring that nodes with the most PE resources are allocated first (line 3). Instances are then deployed onto VMs in the queue; if there exists a VM with idle resources greater than the instance's requirements, the instance is deployed onto it. If the allocation of the instance is successful, the service is also considered successfully deployed (lines 4-9). Finally, global resources and the priority queue are updated accordingly (lines 8-14).

Additionally, during simulation runtime, if the occupancy of VM resources exceeds predefined thresholds, indicating that the virtual machine is overloaded, CloudNativeSim will initiate instance migration. These instances will be migrated from overloaded machines to those with lower loads, carefully considering the resource requirements and dependencies of the services involved. Upon completion of migration, instances will operate on the new virtual machine, and resource statuses will be updated accordingly.

The implementation of the service allocation policy ensures a reliable system state, with a mutual mapping between VMs, instances, and services, allowing resources to be efficiently transferred among them. Additionally, CloudNativeSim maintains extensibility through this method. Users can customize the sorting method of the $sortedQueue$ and the deployment strategy of the instances.

\subsection{Usage History Updating}

To observe the state of a microservice-based system and to guide the simulator's subsequent scheduling activities, a model of resource utilization is essential. In CloudSim, the utilization model serves a similar function, but its inputs only consist of time, and it is dynamically changing based on extensive historical data from real machines, such as those from PlanetLab \cite{planetlab}. While this approach is applicable to cloud-native applications, it lacks sufficient interpretability and struggles with the challenges posed by dynamic environments. Consequently, we propose a more dynamic method that updates the Usage History to record the resource status within the cloud-native application during simulation runtime.

Different from virtual machines, in a cloud-native environment, instances consume minimal resources when idle. As tasks shift from virtual machines to instances, we observe that an instance's resource usage history correlates with its role in processing requests. For example, frequently called instances may consume more bandwidth, while those processing large-scale tasks require more memory and CPU power. In summary, resource usage directly correlates with the number and scale of tasks when active.

Based on the above observations, we attempt to quantify resource usage by the number and duration of cloudlets assigned to an instance for simulation. To address this, CloudNativeSim employs a linear approach to correlate the performance of instances with cloudlets' resource usage. Using the specific calculation method, a cloudlet's execution time depends on the CPU time allocated to it and the processing speed of the instance executing it. The execution time of cloudlets ultimately affects the calculation of response time. Similarly, the memory usage and bandwidth consumption of cloudlets are calculated using a similar approach. Once the resource consumption by the cloudlets is calculated, the corresponding instance's usage data is updated and recorded in the usage history.

Compared to previous simulators, such as CloudSim, which are based on static methods, our calculation method offers a high degree of dynamism, making it highly beneficial for capturing the internal state changes of cloud-native applications. Additionally, by adjusting minimal parameter settings, this simple approach can achieve an excellent fit. This enhances the effectiveness and flexibility of CloudNativeSim in simulating dynamic cloud environments.


\subsection{Service Scaling Policy}\label{sec:scaling}
In the simulation of microservice-based architectures, the service scaling policy is a core function. It can leverage historical resource utilization data to maintain optimal performance, while preventing both resource wastage and inefficiency \cite{Baarzi2021SHOWARRA}. During the operation of the simulator, a service scaling event is triggered at regular intervals to automatically check whether the current service requires scaling, such as when an instance's CPU utilization consistently remains high \cite{wang2022deepscaling}. Once scaling is deemed necessary, the user-defined scaling policy is executed. In CloudNativeSim, we provide the most common horizontal and vertical scaling solutions, as outlined in Algorithms \ref{al:horizontal} and \ref{al:vertical}.

\begin{algorithm}
\caption{Horizontal Service Scaling Algorithm}
\label{al:horizontal}
\hspace*{0.02in} {\bf Input:} 
$scalingList$ (list of replica sets to be scaled), $service$ (the service to be scaled)\\
\hspace*{0.02in} {\bf Output:} 
$result$ (the results of scaling)
\begin{algorithmic}[1]

\State Initialize $result \gets$ false;
\State Initialize $scalingList$ with replications;
\For{$replicaSet$ \textbf{in} $scalingList$} 
    \State Initialize $newReplica \gets$ create a new replica from $replicaSet$;
    \If{success to allocate ($newReplica$)} 
        \State Remove the original $replicaSet$ from $scalingList$;
        \State Bind the new replica to the $service$;
        \State $result \gets$ true;
    \Else 
        \State Delete $newReplica$ to undo the creation of the replication;
    \EndIf
\EndFor
\State \Return $result$

\end{algorithmic}
\end{algorithm}

For horizontal scaling, the simulator operates on a replica set of instances, attempting to replicate and allocate virtual machine resources to each new copy. If resource allocation is successful, the new replica is bound to the service and removed from the scaling list. If the allocation fails, the replica is deleted to avoid resource waste.

The policy for vertical scaling is slightly different. The operation targets the instances needing scaling, calculates each instance's resource requests, and releases the current resources. It then attempts to allocate virtual machine resources for the new resource requests. If resource allocation is successful, the new instance is bound to the service, and the original instance is removed from the scaling list. If the allocation fails, the original instance is restored and added to the failure list for later processing.

\begin{algorithm}
\caption{Vertical Service Scaling Algorithm}
\label{al:vertical}
\hspace*{0.02in} {\bf Input:} 
$scalingList$ (list of instances to be scaled), $service$ (the service to be scaled)\\
\hspace*{0.02in} {\bf Output:} 
$result$ (the result of scaling)
\begin{algorithmic}[1]

\State Initialize $result \gets$ false;
\State Initialize $failedList \gets$ empty list;
\State Initialize $scalingList$ with instances;
\For{$instance$ \textbf{in} $scalingList$}
    \State Initialize $resourceRequests \gets$ $instance$.computeInstanceRequests();
    \State Deallocate resources from $instance$;
    \If{success to allocate ($instance$, $resourceRequests$)}
        \State Remove the original $instance$ from $scalingList$;
        \State Bind the scaled instance to the $service$;
        \State $result \gets$ true;
    \Else
        \State Restore $instance$ to its original state;
        \State Add $instance$ to $failedList$;
    \EndIf
\EndFor
\State \Return $result$

\end{algorithmic}
\end{algorithm}

Additionally, CloudNativeSim employs a threshold-based monitoring system that oversees both resource underutilization and overload scenarios. These thresholds are defined as ranges that trigger scaling decisions: scaling up occurs when resource utilization exceeds upper thresholds to maintain performance, while scaling down is initiated when utilization falls below lower thresholds to optimize resource efficiency.

Utilizing scaling policies greatly enhances the dynamism of CloudNativeSim, enabling it to flexibly respond to changes in workload, ensuring high service availability and efficient resource utilization. In Section \ref{eva:scaling}, we will discuss the impacts of different scaling policies. Users can also customize auto-scaling policies, such as those based on response time or SLO violation rates \cite{mirhosseini2021parslo}, workload levels \cite{auto-scale-by-workload}, and may even implement hybrid policies \cite{xu2022coscal,pbscale}.


In summary, CloudNativeSim implements three key scheduling policies: service placement, usage history tracking, and service scaling. These policies work together to provide a comprehensive solution for simulating cloud-native applications, with particular emphasis on resource efficiency and system adaptability. The extensible design allows users to customize these policies according to their specific requirements while maintaining the core functionality of the simulator.

\section{Verification and Performance Evaluations}\label{sec 6}

In this section, we conducted several experiments on CloudNativeSim to evaluate its simulation performance and verify the effectiveness of the design schemes aforementioned. We will evaluate its practicality from following experiments: (i) the tests of simulation capacity, (ii) the verification of request generator, (iii) the evaluation of response time simulation accuracy compared with realistic testbed, and (iv) the demonstration of different scaling policies.

\subsection{Tests of Simulation Capacity}\label{eva:capacity}

To test the simulation capacity of CloudNativeSim, we conducted a capacity test on a common PC with a 16-core, 2.2 GHz CPU, and limited the running memory to 4 GB. To evaluate the simulator's extreme capacity, we considered the following parameters: the number of services, the total number of instances, the number of requests, and the number of cloudlets. VDifferent parameter configurations resulted varying running time on the realistic testbed (in seconds).

To ensure the reliability of our results, we designed 4 cases and tested 2 parameter sets for each case, conducting 5 repetitive tests for each set. Table \ref{tb:capacity} presents the average running time for these tests:
\begin{table}[ht]
    \renewcommand{\arraystretch}{1.3}
    \centering
    \begin{tabular}{c|cccc|cr}
    \toprule
    Case & Number of Requests & Number of Services & Number of Instances & Number of Cloudlets & Running Time \\
    \midrule
    \multirow{2}{*}{1} & $10^5$ & $1$ & $10^3$ & $10^5$ & 1.95 seconds\\
           & $10^6$ & $1$ & $10^3$ & $10^6$ & 13.29 seconds\\
    \midrule
    \multirow{2}{*}{2} & $10^3$ & $5\times10^3$ & $1$ & $5\times10^6$ & 0.84 seconds\\
           & $10^3$ & $5\times10^4$ & $1$ & $5\times10^7$ & 2.73 seconds\\
    \midrule
    \multirow{2}{*}{3} & $10^4$ & $10^2$ & $3\times10^2$ & $10^6$ & 0.94 seconds\\
           & $10^4$ & $10^3$ & $3\times10^3$ & $10^7$ & 1.40 seconds\\
    \midrule
    \multirow{2}{*}{4} & $10^3$ & $5\times10^3$ & $1.5\times10^4$ & $5\times10^6$ & 4.58 seconds\\
           & $10^4$ & $5\times10^3$ & $1.5\times10^4$ & $5\times10^7$ & 9.56 seconds\\
    \bottomrule
    \end{tabular}
    \caption{Running time for CloudNativeSim capacity test under various parameter sets.}
    \label{tb:capacity}
    \vspace{-1em}
\end{table}

To avoid randomness, we conducted ablation experiments in Case 1 and Case 2, selecting parameter sets with running time under 15 seconds for analysis. In Case 1, we limited the number of services to 1 and the number of instances to one thousand. Under these conditions, CloudNativeSim can simulate up to \(10^6\) requests. In Case 2, we fixed the number of instances at 1 and the number of requests at \(10^4\), demonstrating that CloudNativeSim can handle over \(5 \times 10^4\) services. Notably, the number of cloudlets also increases with the number of requests and services, leading to a substantial rise in the simulator's internal running time.

For typical scenarios, we selected more common quantities in Case 3 and Case 4, maintaining a ratio of 1:3 between number of services and instances. By fixing the number of requests and instances and gradually increasing the number of other objects, both cases showed that CloudNativeSim is capable of simulating over \(10^7\) objects while keeping the running time within 10 seconds.

These test results indicate that CloudNativeSim exhibits considerable elasticity and scalability under various loads. Although the running time significantly increases under high load conditions, it is important to note that the improvement in simulation time compared to physical system testing is substantial. This enhancement allows for more efficient testing and modeling of cloud-native applications, making CloudNativeSim a valuable tool for developers and researchers working with complex microservice architectures.

\subsection{Verification of Request Generation}\label{eva:generator}

\begin{figure}[htbp]
\centering
\begin{subfigure}[b]{0.32\linewidth}
\includegraphics[width=\linewidth]{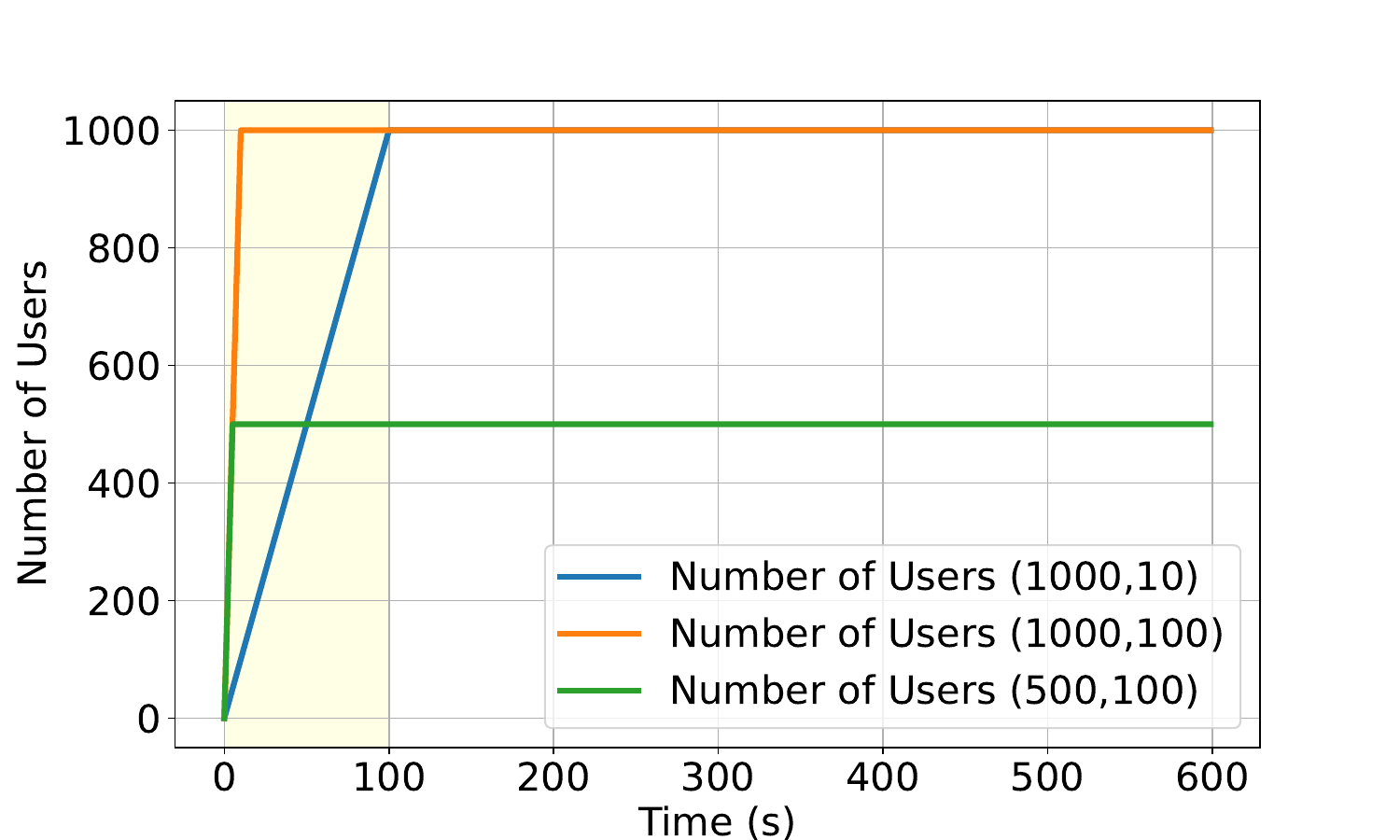} 
\caption{Number of clients over time}
\label{fig:clients}
\end{subfigure}
\hspace{2pt}
\begin{subfigure}[b]{0.32\linewidth} 
\includegraphics[width=\linewidth]{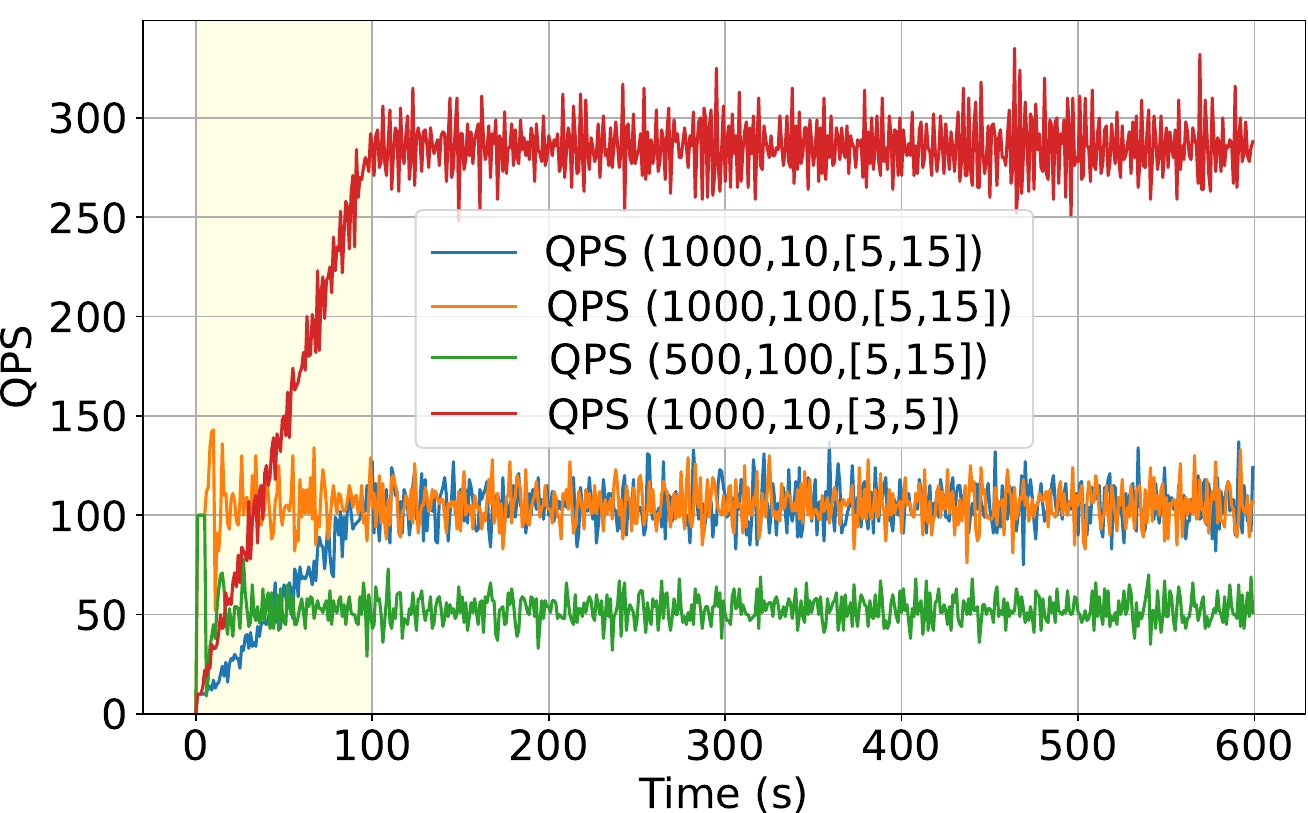}
\caption{QPS over Time}
\label{fig:qps}
\vspace{-0.07cm}
\end{subfigure}
\hspace{2pt}
\begin{subfigure}[b]{0.32\linewidth} 
\includegraphics[width=\linewidth]{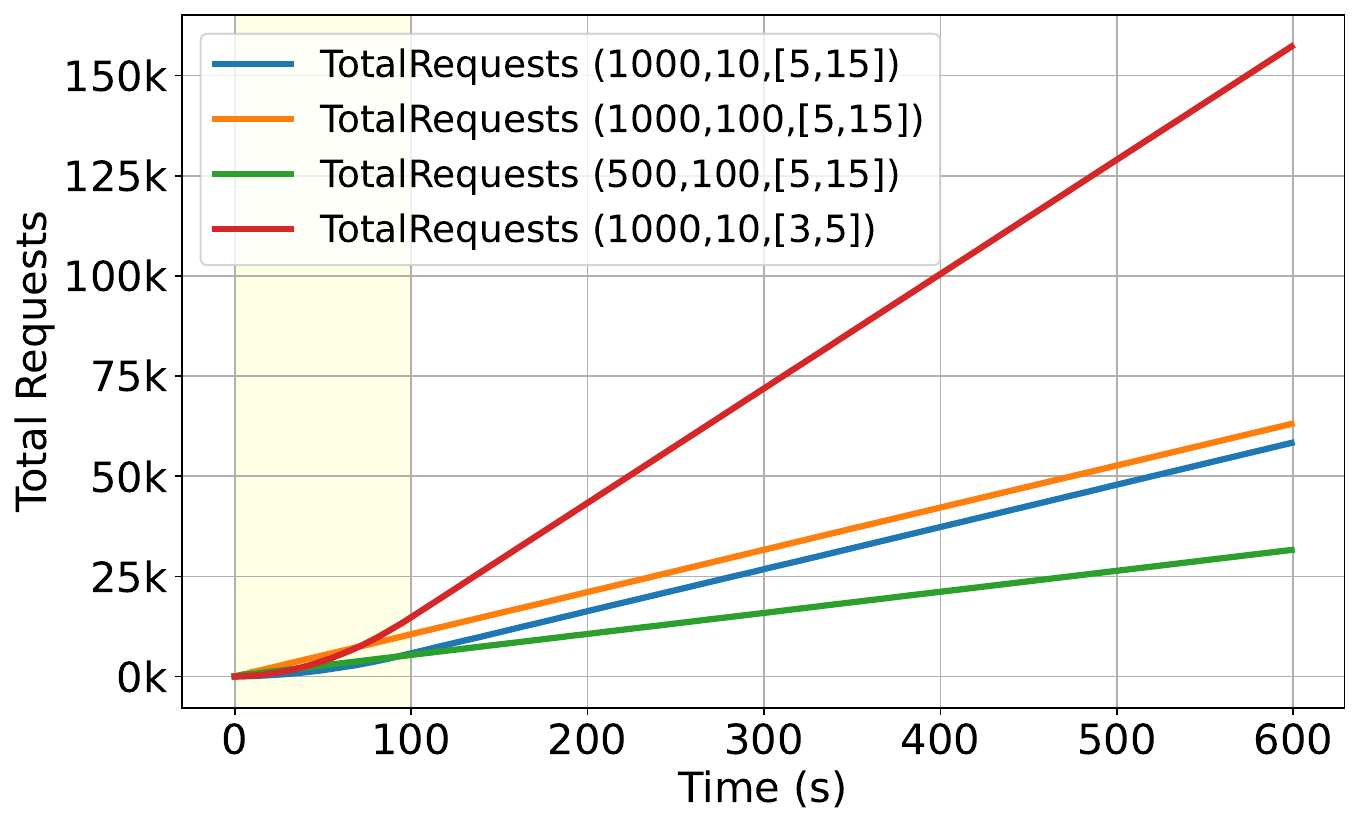} 
\caption{Total number of requests over the simulation period}
\label{fig:requestsNum}
\end{subfigure}

\caption{Simulation results of request generation.}
\label{fig:generator-test}
\vspace{-1em}
\end{figure}

To validate the performance of the Request Generator and the correctness of the modeling in Section \ref{sec:generator}, we select the most important parameters \( (N_c, v, p) \) to conduct four sets of test experiments on the Request Generator. In Section \ref{sec:generator}, Request Generation Modeling, the parameters \( (N_c, v, p) \) represent the total number of clients, the client growth rate, and the request inter-arrival time region, respectively. These parameters are crucial in determining the QPS and the evolution of the total number of clients within the simulation. The total number of clients \( (N_c) \) defines the upper limit of clients that can generate requests, directly impacting the load on the system. The client growth rate \( (v) \) dictates how quickly new clients are added, influencing the rate at which the system load increases. The request inter-arrival time region \( (p) \) determines the frequency of requests from each client, affecting the overall request rate. For example, QPS \((1000, 100, [5,15])\) indicates a configuration with a maximum of 1000 clients, increasing 100 new clients per second, where each client generates requests with an inter-arrival time randomly distributed between 5 and 15 seconds. By collecting and plotting the growth curves of the number of clients, QPS, and total requests, we aimed to evaluate the performance of the Request Generator. Figure \ref{fig:generator-test} illustrates the results produced by the Request Generator across four different test configurations.

Based on the system capacity limits established in our previous experiments, we designed these test configurations to validate the Request Generator's performance within the proven operational bounds. We selected three parameter configurations focusing on different client growth patterns. The first configuration (1000, 10) represents a gradual scaling scenario, where clients are added at a conservative rate of 10 per second until reaching the previously validated maximum of 1000 clients. The second configuration (1000, 100) maintains the same maximum client count but accelerates the growth rate to 100 clients per second, allowing us to examine how growth velocity affects request patterns while staying within known system limits. The third configuration (500, 100) keeps the rapid growth rate but operates at a more moderate client scale, providing insights into the Request Generator's behavior under different steady-state loads. For request patterns, we tested two inter-arrival time regions: [3, 5] seconds for high-frequency requests and [5, 15] seconds for more moderate request rates. These combinations enable us to thoroughly evaluate the Request Generator's performance across various realistic usage scenarios while operating within the validated system capacity established by our previous capacity testing experiments.

In Figure \ref{fig:generator-test}, the segmented curves in the three subplots exhibit distinct characteristics around 100 seconds, coinciding with the period of customer number growth, $\frac{N_c}{v}$. This period is highlighted due to its significance. During this time, the number of clients grows linearly at the spawn rate $v$, eventually stabilizing at the client number $N_c$. This process remains independent of the waiting interval $p$, which explains the presence of only three curves. This behavior aligns with the description provided in Equation \ref{eqa:N(t)}. Initially, the QPS shows approximately linear growth before reaching an oscillatory convergence state. The speed of convergence is influenced by the spawn rate $v$, while the final converged value is determined by both $N_c$ and $p$. Specifically, the final value is negatively correlated with the value and interval size of $p$ and positively correlated with $N_c$, consistent with the description in Equation \ref{eqa:qps}. In subplot \ref{fig:qps}, the blue and orange curves, which share the same $p$ and $N_c$, converge closely and oscillate around 100, as derived from Equation \ref{eqa:qps}. For the total number of requests, the curve exhibits a nonlinear growth trend in the first segment and a linear growth trend in the second segment, consistent with the description in Equation \ref{eqa:R(t)}.

In summary, the Request Generator performed as expected, validating both its performance and the correctness of our model. The observed behaviors in the subplots confirm the theoretical descriptions provided in the equations, demonstrating the model's accuracy in predicting request metrics. The highlighted period of client growth and the subsequent convergence patterns in QPS provide valuable insights into the characteristics of the APIs.

\subsection{Evaluation of Response Time Simulation} \label{eva:fit}

To show the accuracy of CloudNativeSim to simulate realistic response time, we conduct experiments under both realistic testbed and CloudNativeSim.
The realistic experiment was validated on a physical cluster consisting of 10 nodes, including 3 master nodes and 7 worker nodes. Each master node is equipped with a 32-core CPU and 64 GB of memory. Among the worker nodes, 4 nodes each have a 32-core CPU and 64 GB of memory, while the remaining 3 worker nodes are equipped with a 56-core CPU and 128 GB of memory, a 104-core CPU and 256 GB of memory, and a 64-core CPU and 64 GB of memory, respectively.

The experiment utilized the SockShop \footnote{https://github.com/microservices-demo/microservices-demo} microservice-based application, which serves as a simulated e-commerce website designed to demonstrate the advantages and implementation details of microservice architecture. The application is composed of multiple services: the frontend is implemented using NodeJS, the order processing service is written in Java, and other services are developed using Go. The shipping service employs RabbitMQ for asynchronous message passing, enhancing the system's scalability and responsiveness. For database management, SockShop uses MySQL for relational data and MongoDB for unstructured data.

\begin{figure}[htbp] 
    \centering 
    \includegraphics[width=0.8\textwidth]{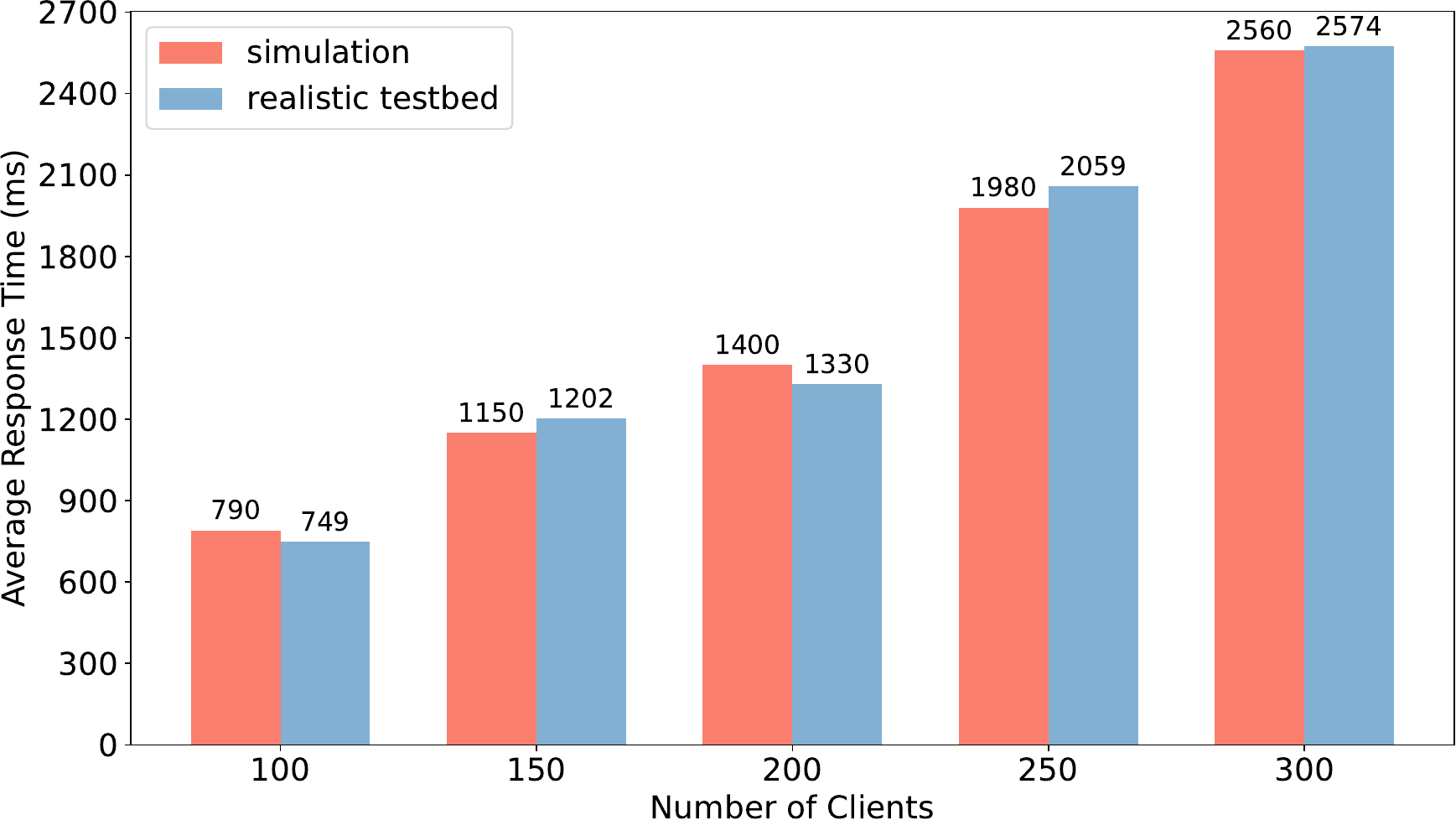} 
    \caption{Comparison of average response time between CloudNativeSim and realistic testbed for the SockShop application.} 
    \label{fig:response}
    \vspace{-1em}
\end{figure}

To assess the simulator's capability in approximating the response time of actual applications, we conducted a stress test on the cluster using Locust, collecting average response time under varying numbers of users as the reference standard. In the Locust script, user wait time were set between 5 to 15 seconds, with a duration limit of 600 seconds. The same settings were applied in the simulator.

Observing the trend in the realistic testbed (blue bars) from the Figure \ref{fig:response}, it is evident that the average response time is positively correlated with the number of clients: as the number of clients increases from 100 to 300, the average response time of the realistic system increases from 749 milliseconds to 2574 milliseconds. This trend is attributable to the increase in QPS and the size of data cached by the system as the number of clients grows.

By adjusting the simulation parameters, the average response time of CloudNativeSim (red bars) increass from 790 milliseconds to 2560 milliseconds. Compared to the realistic testbed under the same conditions, the average simulation accuracy of CloudNativeSim reached a maximum of approximately 99.46\% and a minimum of approximately 94.53\%. This high accuracy is largely due to CloudNativeSim's extensive scalability and dynamic internal mechanisms. These results validate the simulator's effectiveness in simulating the response time of real-world applications.

\subsection{Demonstration of Scaling Policies}\label{eva:scaling}

In realistic testbeds, each application may exhibit different performance characteristics depending on the device and parameter settings. Although we demonstrated in Section \ref{eva:fit} that CloudNativeSim is capable of achieving accuracy, the heterogeneity of cloud-native applications makes it challenging for the simulator to consistently achieve high accuracy for unknown applications. Observing the impact of scaling policies provides another valuable perspective for evaluating the simulator's practicality. This section will examine and analyze the impact of CloudNativeSim using different service scaling policies.

\begin{figure}[htbp] 
\centering
\begin{subfigure}[b]{0.32\textwidth} 
\includegraphics[width=\linewidth]{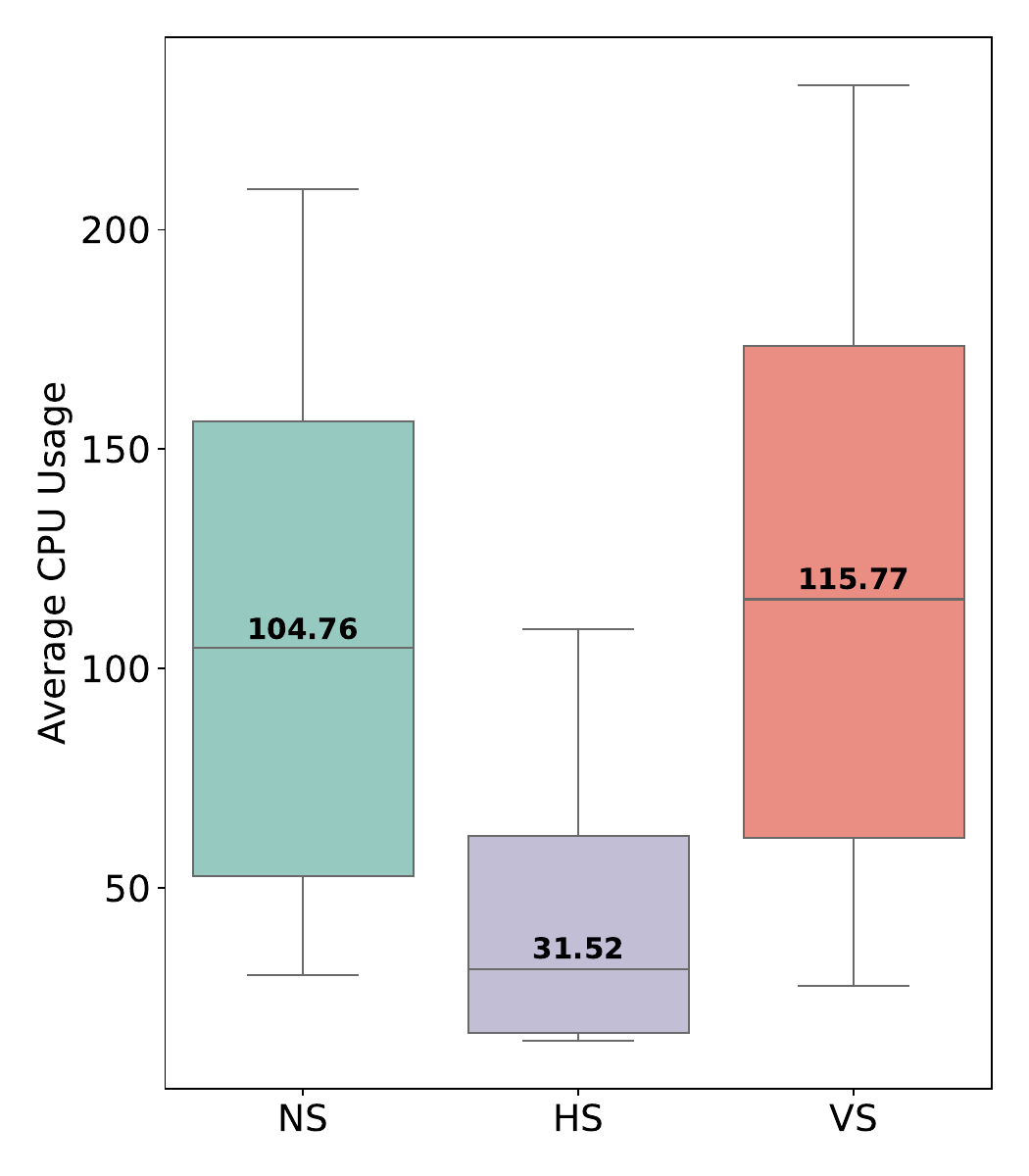} 
\caption{CPU usage under different scaling algorithms with 300 clients}
\label{fig:usage1}
\end{subfigure}
\hspace{0.01\textwidth} 
\begin{subfigure}[b]{0.32\textwidth} 
\includegraphics[width=\linewidth]{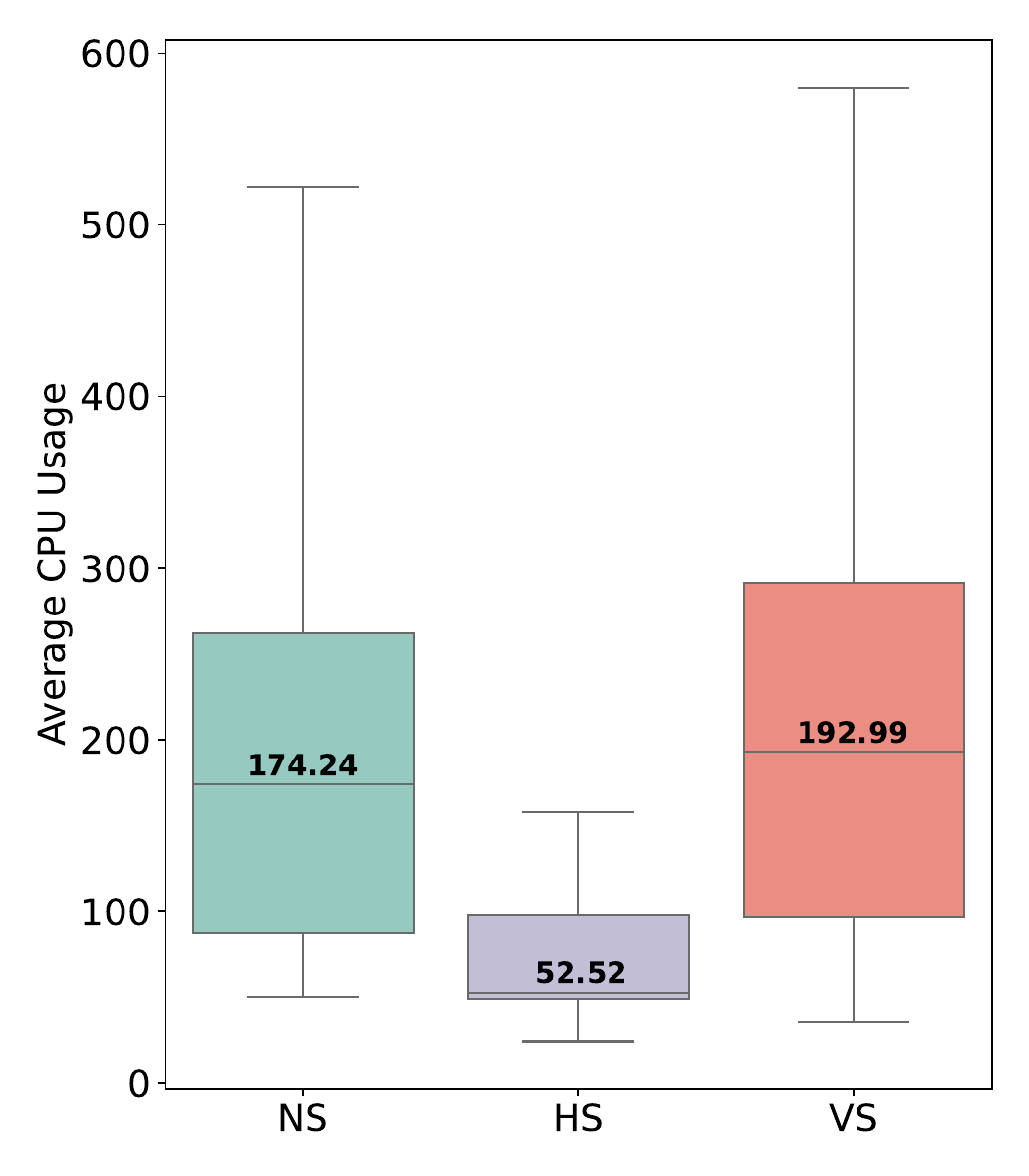} 
\caption{CPU usage under different scaling algorithms with 500 clients}
\label{fig:usage2}
\end{subfigure}
\hspace{0.01\textwidth} 
\begin{subfigure}[b]{0.32\textwidth} 
\includegraphics[width=\linewidth]{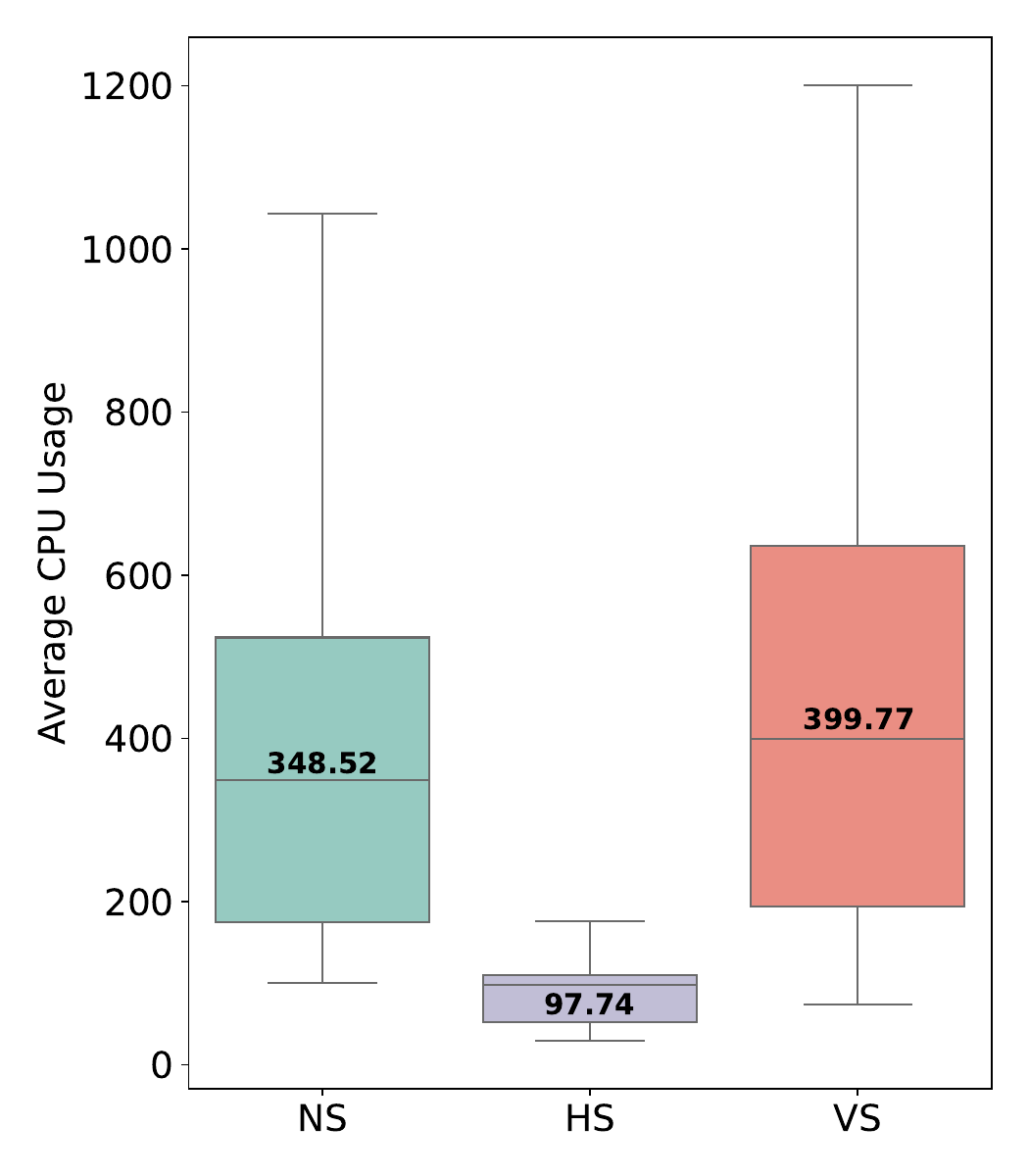} 
\caption{CPU usage under different scaling algorithms with 1000 clients}
\label{fig:usage3}
\end{subfigure}
\caption{Comparison of CPU usage for different scaling algorithms (NS, HS, and VS) under different client loads of 300, 500, and 1000 clients.}
\label{fig:usage exp}
\vspace{-1em}
\end{figure}

CloudNativeSim includes three built-in scaling policies: no scaling algorithm (NS), horizontal scaling algorithm (HS), and vertical scaling algorithm (VS) as introduced in Section \ref{sec:scaling}. In Figure \ref{fig:usage exp}, we test the impact of these three policies on instance usage under different client request loads (300, 500, and 1000 clients) with other configurations consistent with Section \ref{eva:fit}.

Under the given request loads, using NS as a baseline, it is observed that HS maintains relatively low resource usage. Specifically, with 300 clients, HS uses 31.52 milicores compared to NS's 104.76 milicores, which is approximately 69.91\% less. For 500 clients, HS uses 52.52 milicores compared to NS's 174.24 milicores, indicating a reduction of roughly 69.86\%. For 1000 clients, HS uses 97.74 milicores, which is about 71.96\% less than NS's 348.52 milicores. This is because HS scales out by increasing the number of instances (2-4 replicas), providing more options for deploying cloudlets. 

In contrast, VS scales up by increasing the resource allocation limits of individual instances, allowing tasks to consume more resources, thereby resulting in relatively high resource usage. For instance, with 300 clients, VS uses 115.77 milicores, which is about 10.51\% more than NS. With 500 clients, VS uses 192.99 milicores, approximately 10.76\% more than NS. For 1000 clients, VS uses 399.77 milicores, which is about 14.71\% more than NS. Compared to the HS algorithm, VS scales the original instance resources, resulting in significant resource consumption per instance under high load conditions. 

As the load increases, the resource usage of instances also grows up, as seen from the NS results in different figures. This is due to the increasing number of cloudlets that need to be processed simultaneously within instances. Meanwhile, the differences in the effects of the three algorithms become more pronounced. HS scales out by adding more instance replicas to ensure stable service operation, while VS allocates more resources to instances more frequently. 

Overall, these observations highlight the varying efficiency of different scaling policies in managing resource usage under increasing loads, providing insights into the practical applications of CloudNativeSim. The choice between HS and VS depends largely on the specific requirements of the application. For applications that need high performance and can handle short bursts of increased resource consumption, VS is suitable due to its ability to quickly allocate additional resources. Conversely, for applications requiring consistent performance and long-term stability, HS is preferable as it distributes the load across multiple instances, ensuring steady resource usage.

In conclusion, the demonstration validates CloudNativeSim's practicality in employing different scaling strategies demonstrates its robustness and adaptability in simulating various cloud-native application scenarios. By efficiently scheduling resource usage under varying loads, CloudNativeSim proves to be a valuable tool for developers and researchers.

\section{Conclusions, Discussions and Future Work}\label{sec 7}
This paper introduces CloudNativeSim, a toolkit for the modeling and simulation of cloud-native applications. To address the heterogeneity and high dynamism inherent in cloud-native environments, it provides developers and researchers with an efficient modeling solution that can rapidly simulate various scenarios of cloud-native applications locally, thereby reducing risks and costs prior to actual deployment. Furthermore, by implementing different scaling policies for microservice-based architecture, this toolkit presents high scalability, aiding in the in-depth understanding and enhancement of cloud-native application architecture design. Through a series of verification and evaluation experiments, we validate the practicality and reliability of CloudNativeSim.

\subsection{Current Limitations}
Although CloudNativeSim exhibits considerable potential in simulating cloud-native applications and environments, the current version has several inherent limitations:

\begin{itemize}  
    \item \textbf{Limited network traffic modeling}. The current implementation lacks comprehensive network traffic simulation capabilities. While basic communication between microservices is supported, it does not fully capture complex network behaviors such as packet loss and bandwidth constraints in NetworkCloudSim \cite{networkcloudsim}, affecting the accuracy of performance predictions.

    \item \textbf{Lack of edge service mobility support}. The current version does not support the mobility of edge services, crucial for simulating dynamic edge computing scenarios. Unlike iFogSim \cite{ifogsim,ifogsim2}, which effectively handles service mobility, current CloudNativeSim could not simulate scenarios where edge services need to migrate or adapt to moving edge devices.

    \item \textbf{Service discovery mechanism}. The current implementation can be extended to include service discovery features developed in ServiceSim \cite{servicesim}, which would enable more detailed modeling of service communication patterns in dynamic environments, particularly useful for large-scale deployments.

    \item \textbf{Limited fault tolerance simulation}. The current implementation lacks comprehensive fault tolerance capabilities. While basic service failures can be simulated, it does not fully support advanced fault tolerance mechanisms, affecting the simulator's ability to evaluate system reliability and resilience under various failure scenarios.
\end{itemize}

\subsection{Future Work}
Several promising directions can address the current limitations and enhance CloudNativeSim's capabilities:

\begin{itemize}
    \item \textbf{Implementation of service discovery and fault tolerance}. Future work will focus on implementing comprehensive service discovery mechanisms and fault tolerance patterns, enabling more accurate simulation of service interactions in dynamic environments.

    \item \textbf{Enhanced monitoring and observability}. Future versions will incorporate advanced monitoring capabilities, providing deeper insights into system behavior and performance bottlenecks, particularly crucial for large-scale deployments.

    \item \textbf{Optimizing large-scale scenario simulations}. Future work will focus on implementing efficient memory management practices and improved service discovery mechanisms designed for large-scale simulations.

    \item \textbf{Advanced scheduling algorithms}. Future development will incorporate more sophisticated scheduling algorithms to improve resource utilization and overall system performance.

    \item \textbf{Integration with container orchestration platforms}. Future developments will include support for simulating container orchestration platforms like Kubernetes, enabling more realistic modeling of deployment strategies and resource management in cloud-native environments.
\end{itemize}

\section*{Software Availability} CloudNativeSim is now open-sourced on GitHub: \url{https://github.com/CyanStarNight/CloudNativeSim}. The project website also provides various examples and tutorials for user reference.

\section*{Author Contributions}
All authors participated in extensive discussions and contributed to the conceptualization, analysis, and finalization of this manuscript. The specific contributions are as follows: Minxian Xu conceived the original concept and provided overall guidance for the experimental methodology and manuscript development. Jingfeng Wu enhanced the initial concept, implemented the experimental framework, conducted comprehensive experiments, and authored the primary content of the manuscript. Yiyuan He focused on manuscript enhancement through detailed revision, language refinement, and improvement of the overall presentation quality. Kejiang Ye and Chengzhong Xu provided critical insights regarding application scenarios and scalability considerations, while also contributing to the manuscript's refinement.

\section*{Acknowledgments}

\thanks{
This work is supported by Guangdong Basic and Applied Basic Research Foundation (No. 2024A1515010251, 2023B1515130002), Guangdong Special Support Plan (No. 2021TQ06X990), Shenzhen Basic Research Program under grants JCYJ20220818101610023, and JCYJ20240809180935001, Shenzhen Industrial Application Projects of undertaking the National key R \& D Program of China (No. CJGJZD20210408091600002).
}

\bibliography{cloudnativesim}

\end{document}